\documentclass[12pt]{iopart}

\usepackage{graphicx}				
\usepackage{amssymb}
\usepackage{tikz}
\usepackage{cite}
\usepackage{bm}
\usepackage{newcommand_yye}
\usepackage{comment}
\usepackage{import}
\usepackage{subcaption}
\usepackage[colorlinks=true,linkcolor=blue,citecolor=red]{hyperref}
\def\n{\bm{n}}

\def\x{\bm{x}}
\def\X{\bm{X}}
\def\W{\bm{W}}

\def\Var{\mathrm{Var}}

\def\E{\mathbb{E}}

\def\P{\mathbb{P}}
\def\R{\mathbb{R}}

\def\L{\mathcal{L}}

\def\S{\mathcal{S}}
\def\T{\mathcal{T}}

\def\pa{\partial\Omega}
\def\ve{\varepsilon}

\def\erf{\mathrm{erf}}
\def\erfc{\mathrm{erfc}}
\def\erfcx{\mathrm{erfcx}}

\begin{document}

\title[]{Boundary local time on wedges and prefractal curves}

\author{Yilin Ye, Denis S. Grebenkov}

\date{\today}

\address{Laboratoire de Physique de la Mati\`{e}re Condens\'{e}e,  CNRS -- Ecole Polytechnique, Institut Polytechnique de Paris, 91120 Palaiseau, France}
\ead{yilin.ye@polytechnique.edu, denis.grebenkov@polytechnique.edu}

\begin{abstract}
We investigate the boundary local time on polygonal boundaries such as finite generations of the Koch snowflake. 
To reveal the role of angles, we first focus on wedges and obtain the mean boundary local time, its variance, and the asymptotic behavior of its distribution. 
Moreover, we establish the coupled partial differential equations for higher-order moments. 
Next, we propose an efficient multi-scale Monte Carlo approach to simulate the boundary local time, as well as the escape duration and position of the associated reaction event on a polygonal boundary. 
This numerical approach combines the walk-on-spheres algorithm in the bulk with an approximate solution of the escape problem from a sector. 
We apply it to investigate how the statistics of the boundary local time depends on the geometric complexity of the Koch snowflake. 
Eventual applications to diffusion-controlled reactions on partially reactive boundaries, including the asymptotic behavior of the survival probability, are discussed. 
\end{abstract}

%
\vspace{2pc}
\submitto{}
%
%
%

\section{Introduction}
The boundary local time $\ell_t$ plays a significant role in the theory of stochastic processes \cite{Levy, Ito, Freidlin}. For example, the reflected Brownian motion $\X_t$ inside a given Euclidean domain $\Omega \subset \mathbb{R}^d$ with a smooth boundary $\partial \Omega$ satisfies the Skorokhod stochastic equation 
\begin{equation}
\md \X_t = \sqrt{2D} \md \W_t + \n(\X_t) \md \ell_t \,,
\label{eq:xbrownl}
\end{equation}
where $\W_t$ is a standard Wiener process in $\mathbb{R}^d$, $D$ is the constant diffusion coefficient, $\n(\x)$ is the unit normal vector at a boundary point $\x \in \partial \Omega$ oriented inward the domain $\Omega$, and $\ell_t$ is a non-decreasing stochastic process that increments only at encounters of $\X_t$ with the boundary. 
The first term in Eq. (\ref{eq:xbrownl}) represents the ordinary Brownian motion inside $\Omega$, whereas the second term assures its normal reflection back into $\Omega$ after each collision \cite{mckean1975brownian, borodin2015handbook, majumdar2007brownian, Grebenkov20}. 
Curiously, the single stochastic equation (\ref{eq:xbrownl}) determines simultaneously two coupled stochastic processes: the position $\X_t$ and the boundary local time $\ell_t$. In addition, 
the boundary local time can be expressed as 
\begin{equation}
\ell_t = \lim_{\ve \to 0} \frac{D}{\ve} \int\limits_0^t \md {t^\prime} \, \Theta(\ve - \llv \X_{t^\prime} - \partial \Omega \rrv) \,,
\label{eq:elldef}
\end{equation}
where $\llv \x - \partial \Omega \rrv$ is the Euclidean distance between a point $\x$ and the boundary $\partial \Omega$, and $\Theta(x)$ is the Heaviside step function. 
Alternatively, one has 
\begin{equation}
\ell_t = \lim_{\ve\to0} \ve \mathcal{N}_t^{(\ve)} \,,
\label{eq:ellnum}
\end{equation}
where $\mathcal{N}_t^{(\ve)}$ refers to the number of crossings of a thin boundary layer $\partial \Omega_\ve = \{ \x \in \Omega : \llv \x - \partial \Omega \rrv < \ve \}$ of width $\ve$ near $\pa$ 
up to time $t$. 
In this way, the boundary local time $\ell_t$ can be regarded either as the rescaled residence time in a thin boundary layer or as the rescaled number of encounters with the boundary (despite its name, $\ell_t$ has units of length). 

The boundary local time also plays the central role in the encounter-based approach to diffusion-controlled reactions by allowing one to consider general surface reaction or permeation mechanisms \cite{Grebenkov20, Grebenkov21JPA, Grebenkov22a, Bressloff22, Grebenkov22, Bressloff22d, Grebenkov22b, Grebenkov23b, Bressloff23a, Bressloff23b, Bressloff23c}.
There are numerous examples of diffusion-controlled reactions in porous media or on rough catalytic surfaces, including oxygen capture on alveolar surface in the lungs \cite{Felici03, Weibel, Sapoval02, Sapoval21}, heterogeneous catalysis \cite{Filoche08, Bond, Lindenberg, Gheorghiu04, Sapoval01, Filoche05}, electrochemical systems \cite{deLevie65, Pajkossy91, Sapoval93, deLevie90, Halsey92}, and biomolecule recognition \cite{gabdoulline2002biomolecular, basak2019understanding, lyu2023first}. 
As these phenomena are often limited by the diffusive transport of reactants toward a reactive boundary $\pa$, the geometric complexity profoundly influences the encounter statistics, e.g., via the available contact area in a restricted volume. 
A key question arises when the boundary exhibits partial reactivity \cite{Collins49, Berg77, Sano79, Brownstein79, Weiss86, Bressloff08, Grebenkov23f}: how do surface irregularities modulate the reaction kinetics \cite{Grebenkov23f, Galanti16, Sapoval94, Benichou14, Benichou10, Grebenkov18f}? 
As a proxy of the time spent near the boundary or the number of collisions, the boundary local time $\ell_t$ on a microscopically rough or fractal-like boundary can help to reveal the relations between diffusion, geometry, and reactivity, linking stochastic trajectories with macroscopic observables.

\begin{figure}[t!]
  \centering
  \begin{subfigure}{0.45\textwidth}
\includegraphics[trim={0.5cm 5cm 0.5cm 5cm}, clip, width=\textwidth]{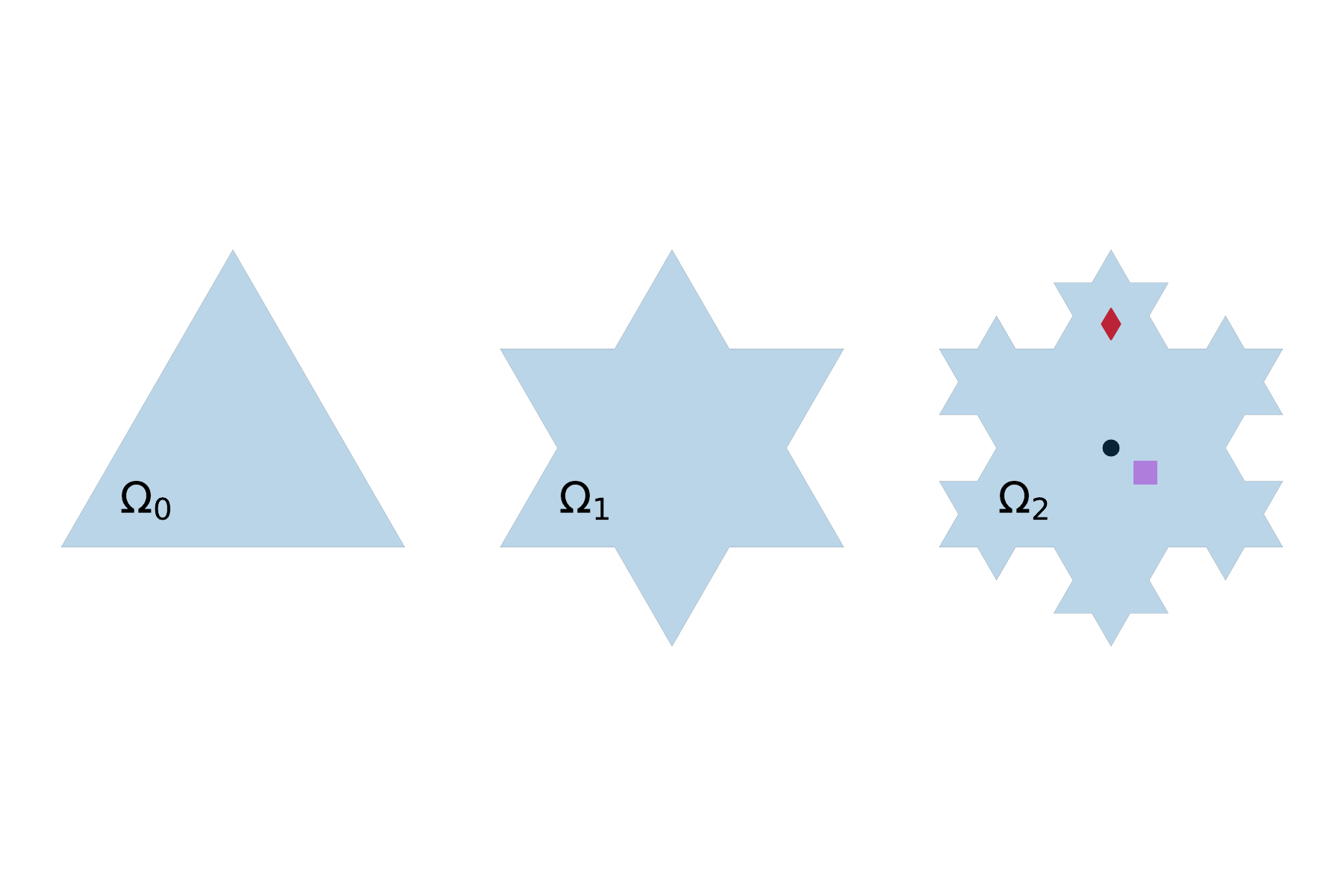}
    \caption{}
    \label{sub:Kochlocal1}
  \end{subfigure}
  \hspace{1.5em} 
  \begin{subfigure}{0.45\textwidth}
\includegraphics[trim={1cm 9.3cm 1cm 9cm}, clip, width=\textwidth]{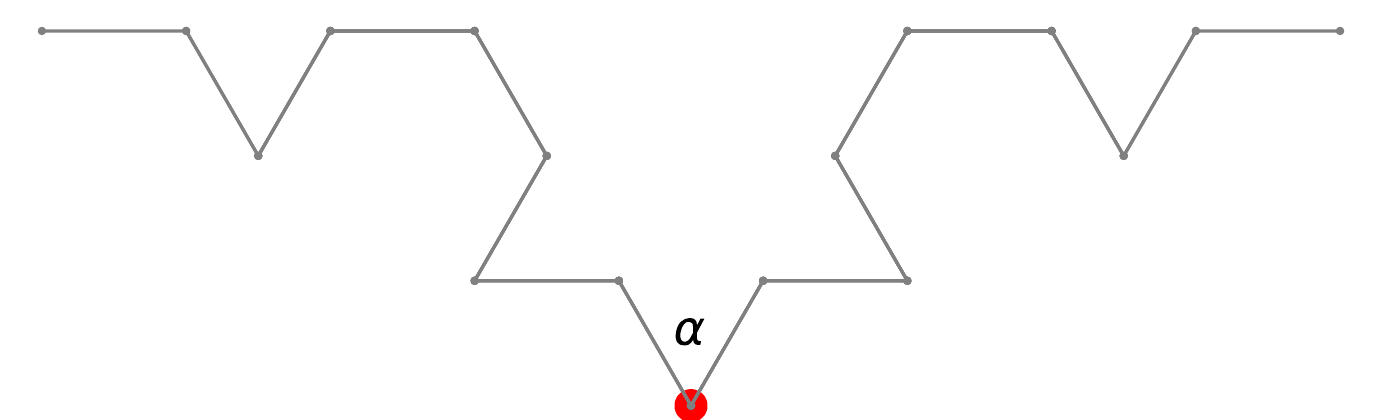}
  \caption{}
    \label{sub:Kochlocal2}
  \end{subfigure}
    \vspace{-0.5em} 
  \caption{\tbf{(a)} Starting from the zeroth generation $g=0$ (an equilateral triangle with the edges of length $L=2$), one constructs finite generations $\Omega_g$ of the Koch snowflake iteratively. In the second generation $\Omg_2$, we indicate two starting points used for validation purposes (see Sec. \ref{sec:valid}): at $(0,5\sqrt{3}/12)$ (red diamond) and at $(1/5,-\sqrt{3}/12)$ (violet square). 
\tbf{(b)} Local environment near the bottom vertex (red dot) of a finite generation of the Koch snowflake of angle $\alpha = \pi/3$.}
  \label{fig:Koch}
\end{figure}

Despite its theoretical significance, the relationship between the distribution of $\ell_t$ and the geometric form of the boundary remains poorly understood. 
Although an analytical solution is available for simple geometries like half-spaces (see below), extensions to wedges and polygonal boundaries confront inherent mathematical challenges \cite{Varadhan85, Williams87, DeBlassie90, Comtet03, Lenzi09, Bras23}. 
For instance, an exact solution for the survival probability in a wedge with partially reactive rays is yet unknown. Even on the numerical side, the most advanced 
algorithms that combine the walk-on-spheres (WOS) method \cite{Muller56} and Skorokhod integral representation \cite{Zhou17, Schumm23} or the escape-from-a-layer approach \cite{ye2024escape}, allow for simulating the reflected Brownian motion and the boundary local time in domains with \tit{smooth} boundaries. 
In turn, undefined or multiple reflection directions near geometric singularities (e.g., corners or cusps) complicate both theoretical formulations and numerical simulations of reflected diffusion in multiscale or self-similar structures like a Koch snowflake (Fig. \ref{sub:Kochlocal1}). 
This gap hinders the analysis of reaction kinetics on rough boundaries which are often modeled as self-similar fractals, 
where the statistics of $\ell_t$ may exhibit geometry-driven anomalies like log-periodic oscillations \cite{Akkermans12, ye2024first}. 
In this paper, we develop the escape-from-a-sector (EFS) approach to bridge this gap and to investigate the boundary local time distributions on prefractal curves. 
We also derive several theoretical results on $\ell_t$ for wedges that will guide our interpretations of numerical results for the Koch snowflake. 

The paper is organized as follows. 
In Sec. \ref{sec:theo}, we analyze the boundary local time distribution in wedges, by focusing on the mean value, the variance, and the asymptotic behavior at large $\ell$ (see also \ref{sec:appPDE}). 
Section \ref{sec:efsa} furnishes the escape-from-a-sector approach for simulating the boundary local time in polygonal domains, and its validation in wedges (Sec. \ref{sec:numwedge}), and in the Koch snowflake (Sec. \ref{sec:valid}). 
In Sec. \ref{sec:num}, numerical simulations are performed to address different aspects: the mean boundary local time for a wedge with two channels (Sec. \ref{sec:recmodel}), the boundary local time distribution in the Koch snowflake (Sec. \ref{sec:numres}), and the local persistence exponent (LPE) of the survival probability in the Koch snowflake (Sec. \ref{sec:LPEk}).
Finally, Sec. \ref{sec:discon} presents further improvements, eventual applications, and conclusions. 
Many technical derivations are relegated to Appendices.

\section{Theoretical results for wedges}
\label{sec:theo}

In this Section, we present our main theoretical results for a wedge of angle $\alpha$: $\Omega = \{ (r,\theta): r>0, 0<\theta<\alpha\}$, written in polar coordinates $(r,\theta)$. Despite the simplicity of this shape, very little is known about the distribution of the boundary local time. 
As shown in \cite{Grebenkov19}, the moment-generating function of $\ell_t$ is given by 
\begin{equation}
S_q(t|\x_0) = \lla e^{-q\ell_t} \rra = \int\limits_0^\infty \md \ell \ e^{-q\ell} \rho_\alpha(\ell,t|\x_0), \quad q > 0 \,,
\label{eq:Sqmg}
\end{equation}
where $\x_0 = (x_0,y_0)$ is the starting point, $\rho_\alpha(\ell,t|\x_0)$ is the probability density of $\ell_t$, and $S_q(t|\x_0)$ is the survival probability of a particle in the presence of partially reactive boundary with the reactivity parameter $q$. The survival probability satisfies:
\begin{align}
\nonumber \partial_t S_q ( t | \x_0) &= D \Delta S_q(t | \x_0) \hskip 10.2mm  \mathrm{in\ } \Omega \,, \\
\nonumber \partial_n S_q + q S_q &= 0 \qquad \qquad \qquad \mrm{on\ } \pa \,, \\
S_q(0|\x_0) &= 1 , \label{eq:survprob} 
\end{align} 
where $\Delta$ is the Laplace operator, and $\partial_n$ is the normal derivative oriented outwards $\Omega$. 
In the case $q = \infty$ (i.e., for the Dirichlet boundary condition on $\pa$), the survival probability $S_\infty(t|\x_0)$ can be found explicitly via separation of variables \cite{Dy08, Chupeau15, LeVot20}. 
In contrast, the Robin boundary condition does not allow for such a separation in polar coordinates, and we are not aware of such results for partially reactive wedges with $0<q<+\infty$. 

Two exceptions are the upper half-plane (the wedge of angle $\pi$) and the positive quadrant (the wedge of angle $\pi/2$), for which the separation of variables can be realized directly in Cartesian coordinates. In the first case, $\ell_t$ is simply the boundary local time of the reflected Brownian motion on the half-line, whose distribution is well known \cite{borodin2015handbook}
\begin{equation}
\rho_\pi(\ell,t|0) = \frac{e^{-\ell^2/(4Dt)}}{\sqrt{\pi Dt}} \,.
\label{eq:rhopi}
\end{equation}
Here and throughout this section, the starting point $\x_0$ is located in the origin, i.e., in the vertex of the wedge, in order to highlight the effect of the corner (see \ref{app:quadrant} for a general location of the starting point). 
In turn, for the positive quadrant, the two rays of the boundary are perpendicular to each other, so that the boundary local time $\ell_t$ is the sum of two independent, identically distributed boundary local times $\ell_t^x$ and $\ell_t^y$ on the horizontal and vertical rays, each obeying the distribution in Eq. (\ref{eq:rhopi}). As a consequence,
\begin{align}  \nonumber
\rho_{\pi/2}(\ell,t|0) &= 
\int\limits_0^\infty \md \ell_1 \, \frac{e^{-\ell_1^2/(4Dt)}}{\sqrt{\pi Dt}}
\int\limits_0^\infty \md \ell_2 \, \frac{e^{-\ell_2^2/(4Dt)}}{\sqrt{\pi Dt}} \delta(\ell_1+\ell_2 - \ell) \\
& = \sqrt{\frac{2}{\pi Dt}} e^{-\ell^2/(8Dt)} \erf \llp \frac{\ell}{\sqrt{8Dt}} \rrp 
\label{eq:rhopi2}
\end{align}
(see \ref{app:quadrant} for an arbitrary starting point). 
Note that the corresponding survival probabilities are 
\begin{align}
S_q(t|0) &= \erfcx(q\sqrt{Dt}) \hskip 12.2mm  (\alpha = \pi), \\
S_q(t|0) &= [\erfcx(q\sqrt{Dt})]^2 \quad (\alpha = \pi/2), 
\end{align}
where $\erfcx(z) = e^{z^2} \erfc(z)$ is the scaled complementary error function. 
For these two cases, it is immediate to compute the moments of the boundary local time; in particular, one gets the mean and the variance 
\begin{align}
\lla \ell_t \rra_0 &= \frac{2}{\alpha} \sqrt{\pi Dt}\,, \label{eq:ellnewmean}
\\ 
\mrm{Var}_0\{\ell_t\} &= \frac{2\pi Dt}{\alpha} \llp 1 - \frac{2}{\pi} \rrp \,, 
\label{eq:ellnewvar}
\end{align}
where the subscript $0$ highlights the starting point at the origin. 
We aim at extending these results to other wedges.

\ssc{Mean value}
We provide two alternative ways to get the mean boundary local time $\lla \ell_t \rra_0$: (i) an intuitive symmetry argument and (ii) a direct computation. In the first approach, we consider the starting point $\x_0$ to be at the vertex. Let us first assume that $\alpha = 2\pi/n$ with an integer $n$ so that the wedge can be replicated $n$ times to cover the whole plane. Due to normal reflections on the boundary of the wedge, random trajectories of ordinary Brownian motion in the plane are statistically equivalent to those of reflected Brownian motion inside the wedge. The boundary local time $\ell_t$ can thus be represented as the sum of $n$ boundary local times $\ell_t^1, \ldots, \ell_t^n$ spent on the rays. As a consequence, the mean value is simply
\begin{equation}
\lla \ell_t \rra_0 = n \lla \ell_t^1 \rra_0 = n \sqrt{\frac{Dt}{\pi}} = \frac{2\pi}{\alpha} \sqrt{\frac{Dt}{\pi}} \,,
\end{equation}
where we used that $\lla \ell_t^1 \rra_0 = \sqrt{Dt/\pi}$. Note that this mean is twice smaller than the mean boundary local time from Eq. (\ref{eq:ellnewmean}) on the whole horizontal line, which can be seen as being composed of two rays. We emphasize that this simple argument does not allow one to get higher-order moments, nor the distribution of $\ell_t$, because $\ell_t^1, \ldots, \ell_t^n$ are not independent. Indeed, if planar Brownian motion spends more time near one ray, it generally spends less time near the other rays. 

In the next step, one can consider $\alpha = 2\pi m/n$ with integer $m$ and $n$. Replicating this wedge $n$ times, one covers the whole plane $m$ times, as if the plane was replicated $m$ times. One can think that Brownian motion switches between $m$ parallel planar ``layers'', thus spending on average $\lla \ell_t^1 \rra_0 / m$ on one ray of one layer. As a consequence, we get again $\lla \ell_t \rra_0 = n \lla \ell_t^1 \rra_0 / m = (2\pi/\alpha) \sqrt{Dt/{\pi}}$. Finally, as ``rational'' angles of the form $2\pi m/n$ are dense in the continuum set of all angles, we conclude that 
\begin{equation}
\lla \ell_t \rra_0 = \frac{2\pi}{\alpha} \sqrt{\frac{Dt}{\pi}} = \frac{2\sqrt{\pi D t}}{\alpha} 
\label{eq:ellrational}
\end{equation}
for any wedge. 

The second method relies on the following representation of the mean boundary local time (see \cite{Grebenkov19} and \ref{sec:appPDE})
\begin{equation}
\langle \ell_t\rangle_{\x_0} = \int\limits_0^t \md t' \, \int\limits_{\pa} \md\x \, D G_0(\x,t'|\x_0),
\label{eq:ellG}
\end{equation}
where $G_0(\x,t|\x_0)$ is the propagator with Neumann boundary
condition. In \ref{sec:appMBLT}, we provide the exact expression for this propagator for any wedge, as well as the evaluation of the double integral in Eq. (\ref{eq:ellG}). This ``brute-force'' computation yields again Eq. (\ref{eq:ellrational}). 
Moreover, we also obtain $\lla \ell_t \rra_{\x_0}$ for any starting point $\x_0$.


\subsection{Variance and higher-order moments}
To access higher-order moments, we establish in \ref{sec:appPDE} a system of coupled partial differential equations between the moments of the boundary local time $\ell_t$. 
Their solution requires the knowledge of the propagator $G_0(\x,t|\x_0)$ that we have already used in Eq. (\ref{eq:ellG}) for evaluating the mean boundary local time. 
Even though this propagator is known explicitly for any wedge (see \ref{sec:app_prop}), the resulting expressions are rather cumbersome even for the second moment of $\ell_t$ (see Eq. (\ref{eq:appB11})). For this reason, we do not further explore this general direction and restrict our attention to the starting point $\x_0$ located at the origin.
In this case, the only length scale of the problem is $\sqrt{Dt}$, and a basic dimensional argument implies that $\lla \ell_t^k \rra_0 = \Lambda_k(\alpha) (Dt)^{k/2}$, where $\Lambda_k(\alpha)$ is (yet unknown) dimensionless prefactor that depends on $k$ and $\alpha$. In particular, Eq. (\ref{eq:ellnewvar}) for the variance, which was valid for $\alpha = \pi$ and $\alpha = \pi/2$, can be generalized as
\begin{equation}
\mrm{Var}_0\{\ell_t\} = v_\alpha \frac{2\pi D t}{\alpha} \llp 1 - \frac{2}{\pi} \rrp \,,
\label{eq:varianceva}
\end{equation}
with a prefactor $v_\alpha$ that depends only on the angle. 
In \ref{sec:appA3}, we compute this prefactor exactly in the case when $\alpha = \pi/n$, with an integer $n$. In particular, we retrieve $v_\pi = v_{\pi/2} = 1$ but show that $v_\alpha$ increases as $\alpha$ decreases. 

\subsection{Asymptotic behavior at large $\ell$} 
We now inspect the asymptotic behavior of the probability density $\rho_\alpha(\ell,t|\x_0)$ of $\ell_t$. 
For a \tit{bounded} domain, the Laplace transform of this probability density admits the following spectral expansion \cite{Grebenkov19}:
\begin{equation}
\tilde{\rho}(\ell,p|\x_0) = \int\limits_0^\infty \md t \ e^{-pt} \rho(\ell,t|\x_0) = \frac{1}{p} \sum\limits_k \mu_k^{(p)} e^{-\mu_k^{(p)}\ell} V_k^{(p)}(\x_0) \int\limits_{\pa} \md\x \, V_k^{(p)}(\x),
\label{eq:rhotilde}
\end{equation}
where $\mu_k^{(p)}$ and $V_k^{(p)}$ are the eigenvalues and eigenfunctions of the generalized Steklov problem:
\begin{align}
\nonumber (p - D \Delta ) V_k^{(p)} &= 0 \hskip 23.8mm  \mrm{in} \quad \Omega \,, \\
\partial_n V_k^{(p)} &= \mu_k^{(p)} V_k^{(p)} \qquad \mrm{on} \quad \pa \,. 
\end{align}
This spectral problem is known to have a discrete spectrum \cite{Levitin}, so that the eigenmodes can be enumerated by a positive integer $k$. 
In turn, as the boundary of a wedge is unbounded, the spectrum of the Steklov problem is not discrete anymore. Nevertheless, for a wedge of angle $\alpha < \pi$, there exists at least one eigenvalue below the essential spectrum (see, e.g., \cite{Levitin08, Khalile18, Pankrashkin23} and references therein), and this eigenvalue determines the asymptotic behavior at large $p$ or, equivalently, at short times. 

To access this behavior, let us inspect the spectrum of the Robin Laplacian in the wedge of
angle $\alpha = 2\Phi$: $\Omega^\prime = \{ (r,\theta)~:~ |\theta|< \Phi\}$ that we rotated by angle $\Phi$ for convenience:
\begin{align}
\nonumber - \Delta u =& \lmd u \hskip 12.3mm  \mrm{in} \quad \Omega^\prime \,, \\
\partial_n u =& \mu u \qquad \mrm{on} \quad \pa^\prime \,,
\end{align} 
with a prescribed parameter $\mu>0$. 
Even though the spectrum is continuous, there may exist a finite number of {\it negative} eigenvalues. In particular, the smallest eigenvalue is $\lambda_1 = - \mu^2/\sin^2(\Phi)$, whereas the associated eigenfunction is $u_1 = \exp\llp -\mu x_0/\sin(\Phi) \rrp$ \cite{Lyalinov21}. 
Setting $\lambda_1 = - p/D$ and employing the duality between the Steklov and Robin problems \cite{Levitin}, one sees that there exists a Steklov eigenpair:
\begin{equation}
\mu_1^{(p)} = \sin(\Phi) \sqrt{p/D}, \qquad V_1^{(p)}(x_0,y_0) = C_1 e^{-\mu_1^{(p)} x_0/\sin(\Phi)} ,
\end{equation}
where $\x_0 = (x_0,y_0)$, and $C_1$ is the normalization constant that ensures
$L^2(\pa)$-normalization of $V_1^{(p)}$:
\begin{equation}
\int\limits_{\pa^\prime} \md\x \, [V_1^{(p)}(\x)]^2 = 1  \qquad \Rightarrow \qquad
C_1^2 = \frac{\mu_1^{(p)}}{\tan(\Phi)} \,. 
\end{equation}
As $\mu_1^{(p)}$ is the smallest eigenvalue, it controls the
asymptotic behavior of Eq. (\ref{eq:rhotilde}) at large $p$. As a consequence, we have as $p\to\infty$
\begin{equation} 
\tilde{\rho}_\alpha(\ell,p|\x_0) \approx \frac{1}{p} \mu_1^{(p)} e^{-\mu_1^{(p)}\ell} V_1^{(p)}(\x_0) \int\limits_{\pa^\prime} \md\x \, V_1^{(p)}(\x) = \frac{2\sin(\Phi)}{\sqrt{pD}} e^{-\sqrt{p/D} \left( x_0 + \ell \sin(\Phi) \right)} \,,  
\end{equation}
from which the inverse Laplace transform yields
\begin{equation}
\rho_\alpha(\ell,t|\x_0) \approx \frac{2\sin(\alpha/2)}{\sqrt{\pi Dt}} e^{-( r_0 \cos(\tta_0 - \alpha/2) + \ell \sin(\alpha/2))^2/(4Dt)} \,, 
\label{eq:asymlong} \end{equation}
where we wrote $x_0$ in terms of the polar coordinates in our conventional wedge $\Omega = \{ (r,\tta): 0 < \tta < \alpha \}$. 
This expression determines the short-time or, equivalently, the
large-$\ell$ behavior.

In contrast, there is no isolated eigenvalue for a wedge of angle $\alpha > \pi$, and the bottom of the essential spectrum is $\sqrt{p/D}$. As a consequence, one may expect the asymptotic behavior $\rho_\alpha(\ell,t|0) \propto e^{-\ell^2/(4Dt)}$ for any $\alpha > \pi$, with eventual power-law corrections.

\ssc{Asymptotic behavior at small $\ell$}

To complete this section, we briefly discuss the asymptotic behavior of the probability density $\rho_\alpha(\ell,t|0)$ at small $\ell$. From a dimensional argument, this is a function of $\ell/\sqrt{Dt}$. According to the explicit solutions in Eqs. (\ref{eq:rhopi}, \ref{eq:rhopi2}), one can expect a power-law behavior,
\begin{equation}  \label{eq:rho_small}
\rho_{\alpha}(\ell,t|0) \propto \frac{1}{\sqrt{Dt}} (\ell/\sqrt{Dt})^{\beta-1}  \quad (\ell\to 0),
\end{equation}
with an exponent $\beta$ that depends only on the angle $\alpha$. In particular, we have $\beta = 1$ for $\alpha = \pi$ and $\beta = 2$ for $\alpha = \pi/2$.
According to Eq. (\ref{eq:Sqmg}), the asymptotic behavior in Eq. (\ref{eq:rho_small}) is tightly related to the large-$q$ asymptotic behavior of the survival probability $S_q(t|0)$ as
\begin{equation}
S_q(t|0) \propto (q\sqrt{Dt})^{-\beta}  \quad (q\to \infty).
\end{equation}
As the survival probability $S_q(t|0)$ is a function of $q\sqrt{Dt}$, its large-$q$ asymptotic behavior is equivalent to the long-time asymptotic behavior.
Since we are not aware of earlier theoretical studies on this quantity for partially reactive wedges, we propose the following heuristic argument. At large $q$ and $t$,
the asymptotic behavior of the survival probability $S_q(t|0)$ is expected to be close to that of $S_\infty(t|r_0)$. Indeed, a particle that managed to survive for
a long time should rapidly move away from the wedge boundary and keep avoiding it. The long-time asymptotic behavior of $S_\infty(t|r_0)$ is well known
(see, e.g., \cite{redner2001guide, metzler2014first, masoliver2018random}): $S_\infty(t|r_0) \propto (r_0/\sqrt{Dt})^{\pi/\alpha}$, so that we conjecture that $\beta = \pi/\alpha$. This conjecture agrees with
the aforementioned exact values of $\beta$ for $\alpha = \pi$ and $\alpha = \pi/2$. 
Further numerical analysis of this conjecture will be reported elsewhere. 

\section{Escape-from-a-sector approach}
\label{sec:efsa}

How can one simulate the boundary local time? 
The simplest method is to discretize the space, such that the reflected Brownian motion is modeled by a random walk on a lattice with a small enough spacing $\ve$, which would require plenty of calculation resources for long trajectories. 
The continuous space simulations can also be realized via the standard walk-on-sphere (WOS) algorithm by Muller \cite{Muller56}, combined with constant displacements inside a thin boundary layer \cite{Zhou17}. 
Moreover, Schumm and Bressloff \cite{Schumm23} implemented the Skorokhod integral representation in planar bounded domains with smooth boundaries.

However, as the boundary of a polygonal domain is not smooth near vertices, we are not aware of Monte Carlo techniques for simulating \tit{efficiently} the boundary local time in such settings. 
To bridge this gap, we develop here the escape-from-a-sector approach based on the spectral decomposition of a suitable escape problem.

We first recall briefly the WOS algorithm for simulating Brownian motion in a confining domain $\Omega \subset \R^2$ with the boundary $\pa$ \cite{Muller56}. 
From a given starting point $\x_0 \in \Omega$, one draws a disk $B_{\rho}(\x_0)$ of radius $\rho = |\x_0 - \pa|$ centered at $\x_0$. 
A continuous trajectory of Brownian motion crosses the boundary of the disk at some random point $\x_1$, which is uniformly distributed on $\partial B_\rho(\x_0)$, at some random escape time $\tau_1$, whose distribution is known explicitly \cite{metzler2014first}. 
As a consequence, a detailed simulation of the Brownian trajectory $\X_t$ inside the disk can be replaced by generating a random escape point $\X_{\tau_1} = \x_1$ at the escape time $\tau_1$. Repeating this procedure, one samples 
a sequence of points $\x_{1}, \x_{2}, \ldots, \x_{k}$ of a random trajectory at random times $t_k = t_{k-1} + \tau_k$, where $\tau_j$ are independent escape times. 
This procedure is iterated until the distance to the boundary becomes smaller than a prescribed threshold $\ve$, e.g., the width of a boundary layer. 
If the boundary is perfectly absorbing, the simulation is stopped, and the current time and position are recorded as the first-passage time to the boundary and its location. 
In our setting, however, the boundary $\pa$ is not perfectly absorbing, so that one needs to simulate reflections on the boundary after the first arrival into the boundary layer. 
This is the most time-consuming step of the former approaches. 

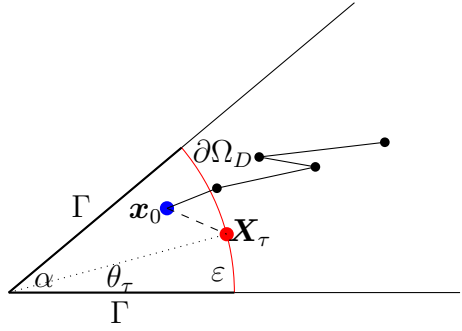
\begin{figure}
\centering
\begin{tikzpicture}[scale=1.0]
	\draw (0 , 0) -- (40 : 6);
	\draw (0 , 0) -- (0 : 6);
	\draw [red] (0: 3) arc [start angle=0, delta angle=40, radius=3];
	\node at (33:3.4) {$\pa_D$};
	\node at (50:1.5) {$\Gamma$};
	\node at (-10:1.5) {$\Gamma$};
	\draw [thick] (0,0) -- (0:3);
	\draw [thick] (0,0) -- (40:3);
	\node at (5:2.8) {$\ve$};
	\node at (7:1.5) {$\theta_\tau$};
	\node at (20:0.5) {$\alpha$};
	\draw [thin, dotted] (0,0) -- (15 : 3);
	\filldraw[color=red, fill=red, thick] (15:3) circle (0.08);
	\node at (30:2.1) {$\x_0$};
	\node at (14:3.3) {$\X_\tau$};
	\filldraw[color=blue, fill=blue, thick] (2.103620006315264, 1.121474541886237) circle (0.08);
	\draw (5, 2) -- (3.3296575021553947, 1.803601058380158);
	\draw (3.3296575021553947, 1.803601058380158) -- (4.076801494398696, 1.6721238500338629);
	\draw (4.076801494398696, 1.6721238500338629) -- (2.767640402581927, 1.3881951724248442);
	\draw (2.767640402581927, 1.3881951724248442) -- (2.103620006315264, 1.121474541886237);
	\draw [dashed] (2.103620006315264, 1.121474541886237) -- (15:3);
	\filldraw[color=black, fill=black, thick] (5,2) circle (0.05);
	\filldraw[color=black, fill=black, thick] (3.3296575021553947, 1.803601058380158) circle (0.05);
	\filldraw[color=black, fill=black, thick] (4.076801494398696, 1.6721238500338629) circle (0.05);
	\filldraw[color=black, fill=black, thick] (2.767640402581927, 1.3881951724248442) circle (0.05);
\end{tikzpicture}
\caption{
Schematic of a random trajectory of a particle moving in a wedge (with black dots representing sampled random positions).
An escape event is initiated when the particle has entered a
circular sector of small radius $\ve$ and angle $\alpha$. The blue dot indicates the current position $\x_0$ of the particle inside the sector, which in local polar coordinates reads as $(r_0,\theta_0)$. The random escape position $\X_\tau = (\ve,\theta_\tau)$ (red dot) is located on the arc $\partial\Omega_D$. The boundary of the circular sector is composed of two reflecting segments $\Gamma$ and the absorbing arc $\partial\Omega_D$.}
\label{fig:scheme2}
\end{figure}

To overcome this limitation, one can simulate the escape from that layer by generating the escape time $\tau$, the escape position $\x'$, and the boundary local time $\ell_\tau$ acquired up to the escape event \cite{ye2024escape}. 
These random variables can be generated from their distributions that are known for a flat layer, which can be considered as a local approximation of a smooth boundary. 
This approach provides an accurate framework for simulating the boundary local time in Euclidean domains with smooth boundaries. 
However, it is not applicable near corners. In this section, we propose an alternative solution, which allows us to handle polygonal boundaries.

Let us consider such a situation when the current position of the particle is closer to a corner of angle $\alpha$ than a prescribed threshold $\ve$ (Fig. \ref{fig:scheme2}). 
After diffusing inside the sector of angle $\alpha$ and radius $\ve$, the particle escapes this sector through a random point on the arc $\partial \Omega_D$. An accurate simulation of the Brownian trajectory inside the sector would involve multiple reflections on the segments and could result in accumulated errors. We aim at replacing this time-consuming step by generating a single escape event (we therefore use the name ``escape-from-a-sector'' approach). 
For this purpose, one needs to generate three random variables: 
the escape time $\tau$, the escape position on the arc (characterized by the angle $\tta_\tau$), and the boundary local time $\ell_\tau$ acquired up to $\tau$. Even though their joint distribution can be formally found \cite{Grebenkov23}, its cumbersome form is not suitable for efficient simulations. For this reason, we propose to substitute random realizations of $\tau$, $\tta_\tau$, and $\ell_\tau$ by their mean values. In other words, we update the counters upon the escape event as: 
\begin{equation} 
t_{n+1} = t_n + \E_{\x_0}\{\tau\}, \qquad 
\ell_{n+1} = \ell_n + \E_{\x_0}\{ \ell_\tau\}, \qquad 
\x_{n+1} = (\ve,\E_{\x_0}\{\theta_\tau\}),
\end{equation}
where the escape position $\x_{n+1}$ is given in local polar
coordinates of the sector. 
We can use the following expressions for the mean values that we derive in \ref{sec:derivation}:
\begin{equation}
\E_{\x_0}\{\tau\} = \frac{\ve^2 - r_0^2}{4D} \,,
\end{equation}
\begin{equation}
\E_{\x_0}\{ \ell_\tau\} = \frac{2\ve}{\alpha} - r_0 \frac{\cos(\theta_0) + \cos(\alpha-\theta_0)}{\sin(\alpha)} 
- \frac{2\ve}{\alpha} \sum\limits_{n=1}^\infty \frac{(1 + (-1)^n)}{\nu_n^2-1} \cos(\nu_n\theta_0) (r_0/\ve)^{\nu_n} ,
\end{equation}  
and
\begin{equation}
\E_{\x_0} \{ \theta_\tau \} 
= \frac{\alpha}{2} \biggl(1 - 4 \sum\limits_{n=1}^\infty \frac{1-(-1)^n}{\pi^2 n^2} \cos(\nu_n \theta_0) (r_0/\ve)^{\nu_n}\biggr),
\end{equation}
where $\nu_n = \pi n/\alpha$. Since $r_0 < \ve$, both sums converge
rapidly.

{
When one needs to generate the boundary local time $\ell_t$, the simulation is stopped when the time counter $t_n$ exceeds a prescribed time $t$. 
Repeating simulations $N$ times, one gets the empirical statistics of $\ell_t$, from which its moments and the empirical density (rescaled histogram) can be estimated. 
In turn, the computation of the survival probability $S_q(t|\x_0)$ requires a different stopping condition: 
 a simulated trajectory is stopped when the boundary local time $\ell_t$ exceeds a random threshold $\hat\ell$ with the exponential distribution: $\P\{\hat\ell > \ell\} = e^{-q\ell}$ \cite{Grebenkov20}. 
In other words, the simulation is stopped when $\ell_n > \hat\ell$, and the first-reaction time $\T$ is assigned to be $t_n$. 
Repeating simulations $N$ times, one gets the empirical statistics of $\T$, from which its moments, the empirical density, and the survival probability can be estimated. 
}

\ssc{Implementation}
We launch Monte Carlo simulations of the reflected Brownian motion starting from a fixed point $\x_0$ inside a given domain, e.g., a wedge or a polygon like finite generations of the Koch snowflake. The latter can be constructed iteratively starting from an equilateral triangle of length $L=2$ and replacing each segment by a simple generator (Fig. \ref{sub:Kochlocal1}). 
In this way, the $g$-th generation $\Omg_g$ of the Koch snowflake is a polygonal domain whose boundary is composed of $3 \times 4^g$ segments of length $h_g = L/3^g$. As $g$ increases, one adds finer and finer geometrical details such that the perimeter of the boundary, $3 \times 4^g \cdot h_g$, diverges. The limiting domain $\Omega_\infty$ has a fractal boundary that is characterized by the fractal dimension $d_f = \ln4 / \ln3 \approx 1.26$ \cite{Mandelbrot}.  

\begin{figure}[t!]
\centering
\includegraphics[width=0.7\linewidth]{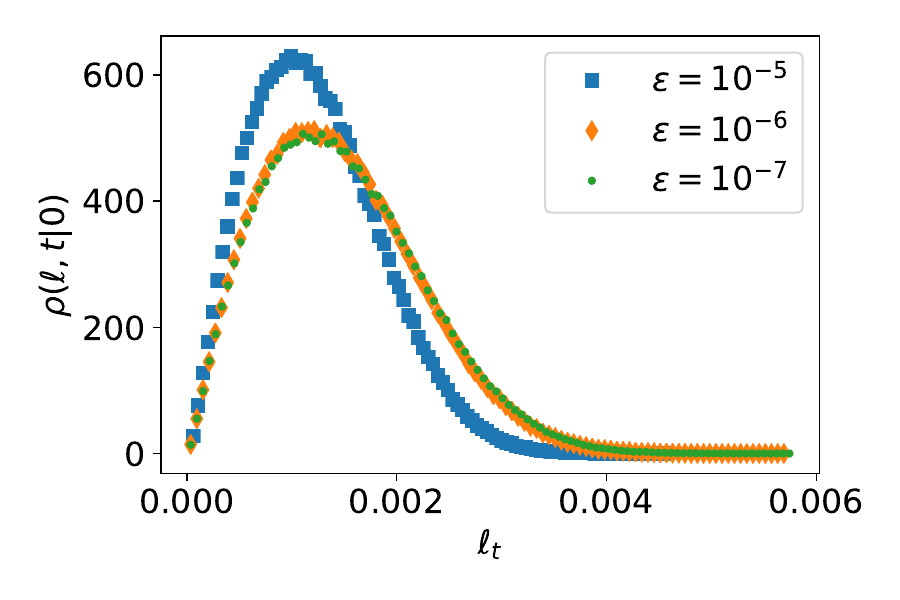} 
\caption{Comparison of empirical PDFs of the boundary local time obtained by Monte Carlo simulations in the Koch snowflake at $g=10$ with $L=2$, $D=1$, $t = 10^{-7}$, and $N = 10^6$ particles. 
The boundary layer thickness is chosen as $\ve = 10^{-5}$ (blue squares), $\ve = 10^{-6}$ (orange diamonds), and $\ve = 10^{-7}$ (green dots). }
\label{fig:histvad}
\end{figure}

A prescribed parameter $\ve$ is introduced as the thickness of the boundary layer. 
When the current position $\x_n$ of the particle is far away from the boundary, i.e., the distance $\rho = \llv \x_n - \pa \rrv$ is larger than $\ve$, the WOS algorithm is executed. 
When $\x_n$ is near the boundary but far away from corners, i.e., $\rho < \ve$ but $\rho_v = \llv \x - \pa_v \rrv > \ve$ where $\pa_v$ is the ensemble of vertices, the EFL method \cite{ye2024escape} with the flat-layer approximation is exploited. If $\x_n$ is near any vertex, i.e., $\rho_v < \ve$, the EFS approach is employed.

The choice of $\ve$ is a compromise between accuracy and rapidity. For an accurate computation, one needs to ensure that $\ve$ is smaller than the smallest geometric feature of the boundary. For instance, for the $g$-th generation of the Koch snowflake, one needs $\ve \ll h_g$. 
Figure \ref{fig:histvad} compares three empirical probability density functions (PDFs) of the boundary local time $\ell_t$ in $\Omg_{10}$, obtained with different values of $\ve$. As $h_{10} \approx 3.4 \times 10^{-5}$, both choices $\ve = 10^{-6}$ and $\ve = 10^{-7}$ are small enough and thus yield almost indistinguishable PDFs. In turn, the larger value $\ve = 10^{-5}$ yields a different (wrong) PDF. According to this brief verification, we use $\ve = 10^{-6}$ for generations $g$ up to 10. 
In turn, $\ve = 10^{-7}$ is taken for higher generation $g = 11, 12$. 
A home-built code was written in Fortran 90 for numerical simulations, while data were analyzed in MATLAB and Python. 

It was checked that the CPU time for a numerical simulation of $\ell_t$ depends on four parameters: the generation $g$, the number of particles $N$, the time $t$, and the boundary layer thickness $\ve$, such that CPU $\propto gNt/\ve$. The dependence of CPU on $g$ is almost linear because we exploit the geometry-adapted fast random walk (GAFRW) algorithm for distance calculations \cite{Grebenkov05b}. 
When the stopping condition concerns the reactivity $q$, we have CPU $\propto gN/(q\ve)$. 
For small reactivity $q$ (almost inert surfaces), a larger number of particles should be taken to diminish statistical fluctuations. 

\subsection{Validation of EFS approach on wedges}
\label{sec:numwedge}

\begin{figure}[t!]
\begin{center}
\includegraphics[width=0.7\linewidth]{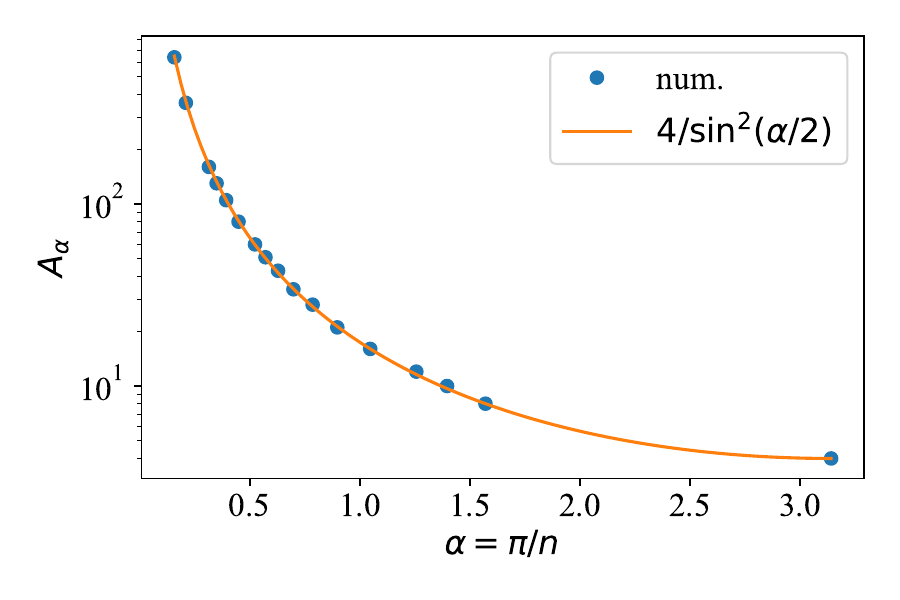}
\end{center}
\caption{
The coefficient $A_\alpha$ of the large-$\ell$ asymptotic behavior of the probability density distribution $\rho_\alpha(\ell,t|0)$ as a function of angle $\alpha$. 
Circles present the numerical fitting via Eq. (\ref{eq:rhoaA}) from the empirical density obtained by Monte Carlo simulations with $N = 10^6$ particles and $Dt = 1$ for each angle, whereas the solid line shows the theoretical prediction (\ref{eq:Aalpha}). The width of the boundary layer is taken as $\ve = 10^{-3}$. 
}
\label{fig:Aalpha} 
\end{figure}

To validate the EFS approach on wedges, we first investigate the statistics of the boundary local time for different wedge angles. For this purpose, we set the starting point at the vertex of the wedge of angle $\alpha$. 
From simulations, we get an excellent agreement with Eq. (\ref{eq:rhopi2}) for $\alpha = \pi/2$ (figure is not shown). 
We also retrieve numerically the Gaussian right tail of the distribution: 
\begin{align}
\rho_\alpha(\ell,t|0) &\propto e^{-\ell^2/(A_\alpha Dt)}  \qquad (\ell\to\infty) \,, \label{eq:rhoaA} \\
A_\alpha &= 4/\sin^2(\alpha/2) \,, 
\label{eq:Aalpha}
\end{align}
for a broad range of angles from $\pi/20$ to $\pi$ (Fig. \ref{fig:Aalpha}), in perfect agreement with Eq. (\ref{eq:asymlong}).

\begin{figure}[t!]
\begin{subfigure}{0.48\textwidth}
\includegraphics[height=60mm]{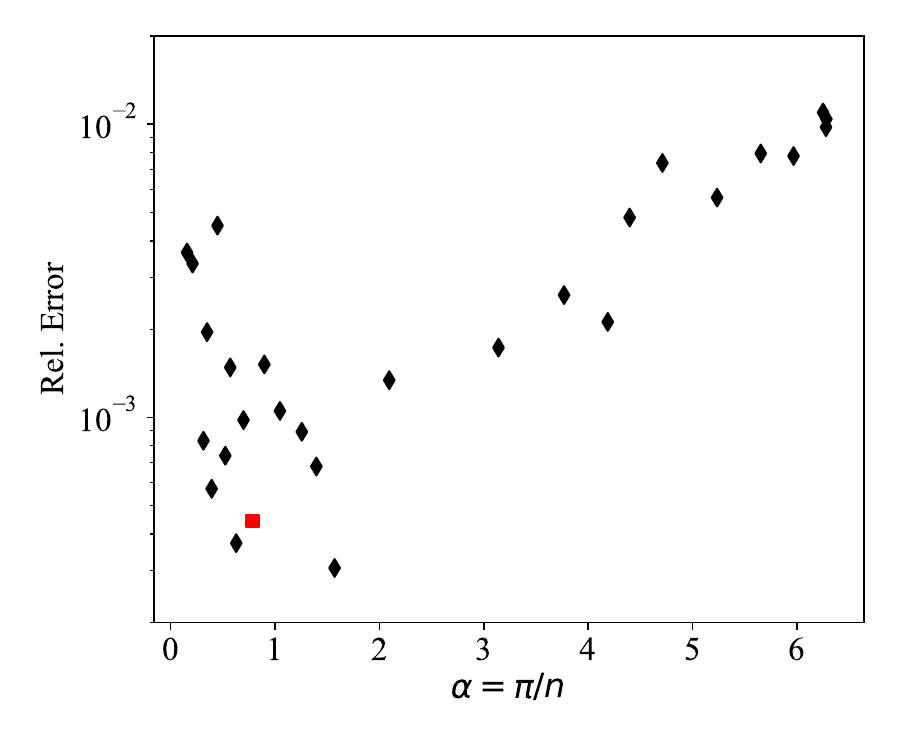}
\caption{}
\label{fig:wedge_meanerr}
\end{subfigure}
\begin{subfigure}{0.48\textwidth}
\includegraphics[height=60mm]{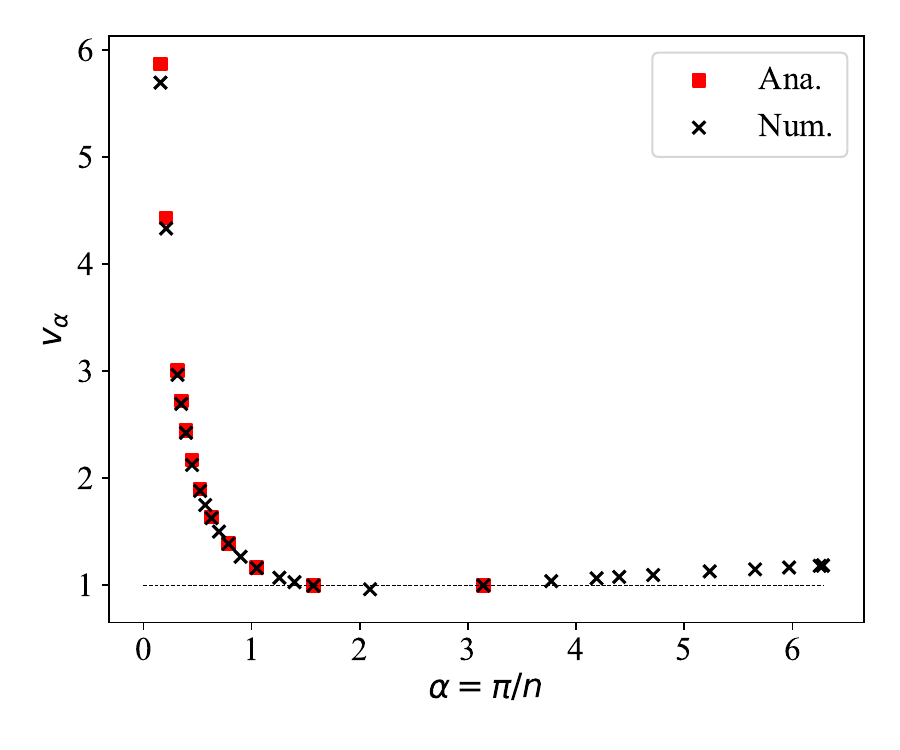} 
\caption{}
\label{fig:wedge_varv}
\end{subfigure}
\caption{
\tbf{(a)} Relative error $\llv \lla \ell_t \rra_0^\mrm{num} / \lla \ell_t \rra_0 - 1 \rrv$ of the mean boundary local time $\langle \ell_t \rangle$ as a function of the angle $\alpha$ of the wedge, with $Dt = 1$. The analytical values are given by Eq. (\ref{eq:ellnewmean}), whereas numerical values were obtained by Monte Carlo simulations with $N = 10^6$ particles and $\ve = 10^{-3}$. Positive values are shown as squares ($\alpha = \pi/4$), while negative values are shown as diamonds. 
\tbf{(b)} The prefactor $v_\alpha$ from Eq. (\ref{eq:varianceva}) as a function of the angle $\alpha$ of the wedge. 
The analytical values (red squares) are given by Eq. (\ref{eq:valpha}). 
Numerical values (black crosses) were obtained by Monte Carlo simulations with $N = 10^6$ particles and $\ve = 10^{-3}$. The thin dashed line presents 1 (the exact value of $v_\alpha$ for $\alpha = \pi$ and $\alpha = \pi/2$). 
}
\label{fig:wedge_var}
\end{figure}

Another validation of the EFS approach employs our theoretical prediction (\ref{eq:ellnewmean}) for a broad range of angles from $\pi/20$ to $2\pi$ (Fig. \ref{fig:wedge_meanerr}). 
The relative error between theoretical and numerical values of $\lla \ell_t \rra_0$ is small and can be attributed to a finite statistics and eventual (minor) errors of our approximate simulation scheme. 
Figure \ref{fig:wedge_varv} shows the prefactor $v_\alpha$ in Eq. (\ref{eq:varianceva}) as a function of the angle $\alpha$ of the wedge. 
For angles of the form $\alpha = \pi/n$, theoretical values obtained in \ref{sec:appA3} are in good agreement with numerical predictions. 
Curiously, the factor $v_\alpha$ exhibits a non-monotonous dependence on $\alpha$ and increases as $\alpha \to 0$. 

\subsection{Validation via spread harmonic measures}
\label{sec:valid}

\begin{figure}[t!]
\centering
\includegraphics[width=0.9\linewidth]{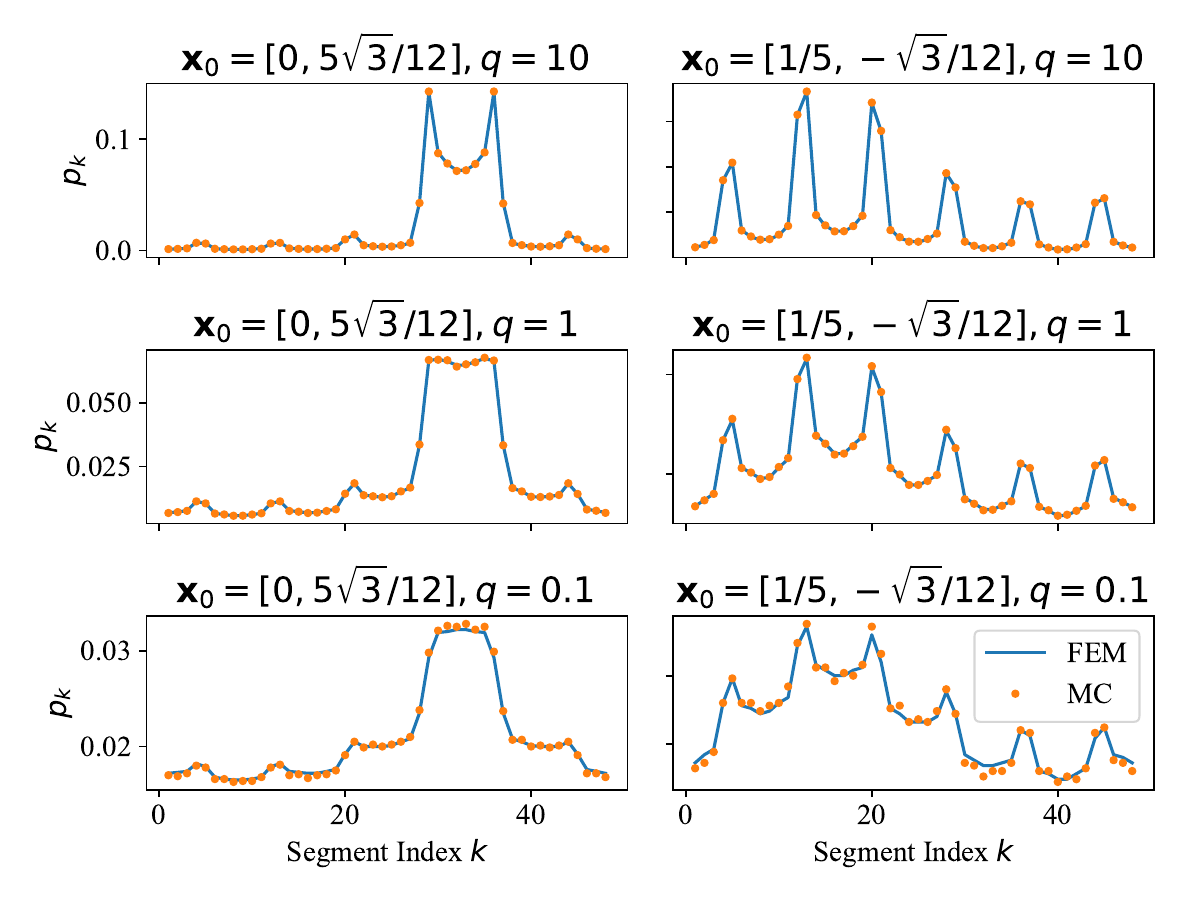}
\caption{Spread harmonic measure on the second generation $\Omg_2$ of the Koch snowflake: comparison between the finite-element method (blue lines) and our Monte Carlo technique (orange dots). Parameters are: the length $L = 2$, the boundary layer width $\ve = 10^{-3}$, particle number $N = 10^6$, reactivity $q \in [10, 1, 0.1]$ for each row from the top to the bottom, and the starting point $\x_0$ is located at $(0,5\sqrt{3}/12)$ for the left column and at $(1/5,-\sqrt{3}/12)$ for the right column. 
The segment index $k$ starts from the left bottom and increases anti-clockwise. 
}
\label{fig:shm}
\end{figure}

For a partially reactive surface, the spread harmonic measure characterizes the spatial distribution of successful reaction events \cite{Grebenkov06, Grebenkov15}. 
In order to validate the accuracy of the proposed EFS method in polygonal domains, we compute numerically the spread harmonic measure for several generations of the Koch snowflake. More precisely, we compute the probability $p_k$ of reacting on the $k$-th segment of the boundary of $\Omg_g$.  
If the starting point is located at the center of the equilateral triangle $\Omega_0$ or the first generation of the Koch snowflake $\Omega_1$, the the symmetry implies equal probabilities $p_k$ for all segments, regardless of the reactivity $q$. We checked this statement numerically (results are not shown).   
Besides, we examine starting points out of the center of the second generation $\Omega_2$, namely, $(0,5\sqrt{3}/12)$ and $(1/5,-\sqrt{3}/12)$, as shown in Fig. \ref{sub:Kochlocal2}. 
As a benchmark, an implementation of a finite-element method (FEM) was realized, where the spectral expansion of the spread harmonic measure was truncated (see \cite{Grebenkov20}) to 30 terms, and the maximal mesh size was 0.005 \cite{chaigneau2024numerical}. The accuracy of this FEM computation was checked by varying the truncation order and the mesh size. 
In Monte Carlo simulations, we set $\ve = 10^{-3} \ll h_2 \approx 2.2 \times 10^{-1}$ to ensure an accurate estimate of the boundary local time.  

Figure \ref{fig:shm} shows the probabilities $\{ p_k \}$  obtained by FEM (blue lines) and Monte Carlo (MC) techniques (orange dots). 
Since the first point $(0,5\sqrt{3}/12)$ is located on the vertical axis of symmetry, we observe the symmetrical pattern of $\{ p_k \}$ in three panels in the left column. 
For the second point $(1/5,-\sqrt{3}/12)$, the distribution exhibits six peaks as expected. 
With the decrease of reactivity, from top to bottom, we observe fewer differences in $\{ p_k \}$ among segments. Both methods provide almost identical results that demonstrate the accuracy and capacity of the EFS approach.


\section{Numerical results}
\label{sec:num}

In this Section, we present our main numerical results for finite generations $\Omega_g$ of the Koch snowflake, when the starting point is located at one of the vertices. As the zeroth generation $\Omega_0$ is just an equilateral triangle of side $L$, the distribution of the boundary local time in $\Omega_0$ is expected to be close to $\rho_{\pi/3}(\ell,t|0)$ in the wedge of angle $\pi/3$, when $t$ is much smaller than $L^2/D$. Indeed, when $t \ll L^2/D$, only a few trajectories of the reflected Brownian motion can reach the opposite side of the triangle and thus "feel" the difference between the triangle and the wedge. At the next iteration, the edges of the equilateral triangle are "decorated" by small triangles. Is this modification of the boundary "beneficial", i.e., does it enhance or diminish the average number of encounters with the boundary? How does it reshape the distribution of $\ell_t$? Answering these basic questions is actually not simple. On the one hand, the perimeter of the boundary, which is typically accessible to the particle up to time $t$, is increased, that may increase $\langle \ell_t\rangle$. On the other hand, two straight segments of the triangle, which were easily accessible to the diffusing particle, are now removed, while the added longer segments are less accessible due to diffusion screening (see \cite{Traytak92, Sapoval94, Felici03, Grebenkov05, Andrade07, Filoche08} and references therein). The iterative construction of the next generations makes the answers even more difficult. For this reason, we first consider a simple setting of a wedge with two channels (Sec. \ref{sec:recmodel}) and then proceed to finite generations of the Koch snowflake (Sec. \ref{sec:numres}). 

\subsection{Wedge with two channels}
\label{sec:recmodel}

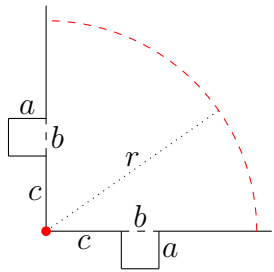
\begin{figure}[t!]
\centering
\begin{tikzpicture}[scale=1.0]
	\draw (1.5,0) -- (3,0);
	\draw (0,1.5) -- (0,3);
	\draw (0,0) -- (1,0);
	\draw (0,0) -- (0,1);
	\draw [dashed] (1,0) -- (1.5,0);
	\draw [dashed] (0,1) -- (0,1.5);
	\draw (1,0) -- (1,-0.5);
	\draw (1,-0.5) -- (1.5,-0.5);
	\draw (1.5,-0.5) -- (1.5,0);
	\draw (0,1) -- (-0.5,1);
	\draw (-0.5,1) -- (-0.5,1.5);
	\draw (-0.5,1.5) -- (0,1.5);
	\draw [dashed, red] (0: 2.8) arc [start angle=0, delta angle=90, radius=2.8];
	\draw [dotted] (0,0) -- (35:2.8);
	\filldraw[color=red, fill=red, thick] (0, 0) circle (0.05);
	\node at (40:1.5) {$r$};
	\node at (0.5,-0.15) {$c$};
	\node at (-0.15,0.5) {$c$};
	\node at (1.25,0.2) {$b$};
	\node at (0.15,1.25) {$b$};
	\node at (1.65,-0.25) {$a$};
	\node at (-0.25,1.65) {$a$};
\end{tikzpicture}
\caption{
Schematic of a wedge of angle $\alpha = \pi/2$ with two additional rectangular channels on each ray. 
The starting point $\x_0$ is located at the vertex of the wedge (red dot). 
}
\label{fig:recpi2}
\end{figure}

Before proceeding to the analysis of the boundary local time in self-similar polygonal domains, we consider a minor alteration of the wedge by adding two identical rectangular channels on both rays (Fig. \ref{fig:recpi2}). As in the case of the first generation of the Koch snowflake, 
it is not clear a priori whether such a modification would increase or decrease the mean boundary local time. 
We investigate here $\lla \ell_t \rra_0$ as a function of the parameters $a$, $b$, and $c$ of the channels (here we restrict our discussion to the case $a=b$).

\begin{figure}[t!]
\centering
\includegraphics[width=0.7\linewidth]{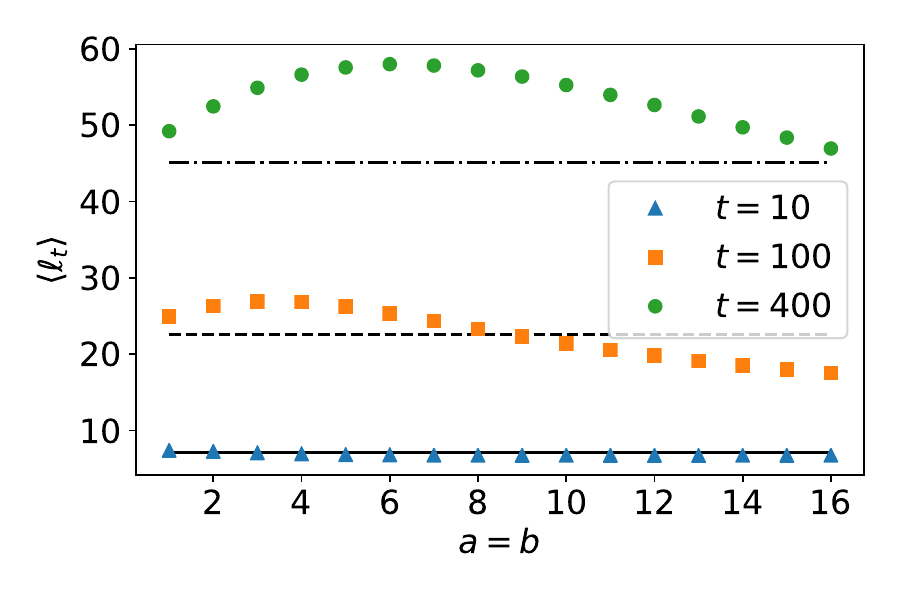} 
\caption{
Mean boundary local time $\lla \ell_t \rra_0$ in a wedge of angle $\pi/2$ with two channels (Fig. \ref{fig:recpi2}) as a function of $a=b$, with $c=5$, $D = 1$, $N = 10^6$ particles and $\ve = 10^{-2}$ for different times: $t=10$ (blue triangle), $t=100$ (orange square), and $t=400$ (green dot). 
Horizontal lines indicated $4 \sqrt{Dt/\pi}$ that corresponds to $\lla \ell_t \rra_0$ in the wedge without channels (i.e., $a=b=0$). 
}
\label{fig:ellm_aeqb}
\end{figure}

Figure \ref{fig:ellm_aeqb} shows the mean boundary local time in the domain with $c=5$ and $a=b$ at $t=10, 100, 400$ (with $D=1$). 
At $t = 10$ (shown by triangles), the diffusion length $\sqrt{Dt} \approx 3.3$ is smaller than $c$, i.e., most trajectories do not reach the channel, and $\lla \ell_t \rra_0$ is very close to that of the wedge of angle $\pi/2$, independently of the channel width $a$. 
At $t = 100$, the diffusion length $\sqrt{Dt} = 10$ is comparable to $c$ so that the channels start to affect the statistics of encounters; in particular, the mean boundary local time increases and then decreases, with the maximum around $a = 3$. 
Moreover, when $a \geq 9$, $\langle \ell_t\rangle_0$ becomes smaller than the mean value for the wedge without channels. In other words, we see that the presence of the channels can either increase or decrease the mean boundary local time, depending on its size: it is beneficial for $a \lesssim \sqrt{Dt}$ and detrimental for $a \gtrsim \sqrt{Dt}$. This is not surprising. In the former case, the larger accessible perimeter tends to increase $\langle \ell_t\rangle_0$. In turn, when $a \gtrsim \sqrt{Dt}$, the particle does access the whole channel, and the accessible perimeter is smaller. 
At $t = 400$, the above observation remains valid, but the maximum is shifted to $a \approx 6$.  
This observation illustrates that the diffusion length scale $\sqrt{Dt}$ is not sufficient to determine the maximum, which depends on the shape of the whole domain. 

\subsection{Statistics of the boundary local time in the Koch snowflake}
\label{sec:numres}

We employ the EFS method to investigate systematically the statistics of the boundary local time in the Koch snowflake, for different times $t$ and generations $g$. 
In this study, we aim at revealing the effect of boundary complexity onto the boundary local time. For this reason, we start all simulations from a vertex of angle $\pi/3$ (see Fig. \ref{sub:Kochlocal2}) and consider a broad range of intermediate times, such that $t \ll L^2/D = 4$. 
In this case, the diffusing particle does not have time to reach the central zone of the Koch snowflake and thus to ``feel'' confinement in a bounded domain. In other words, the particle diffuses near a prefractal boundary, as if the domain was unbounded. In this way, we can reveal the effect of boundary irregularity as compared to the flat boundary of the wedge of angle $\pi/3$.

\begin{figure}[t!]
\centering
\includegraphics[width=0.7\linewidth]{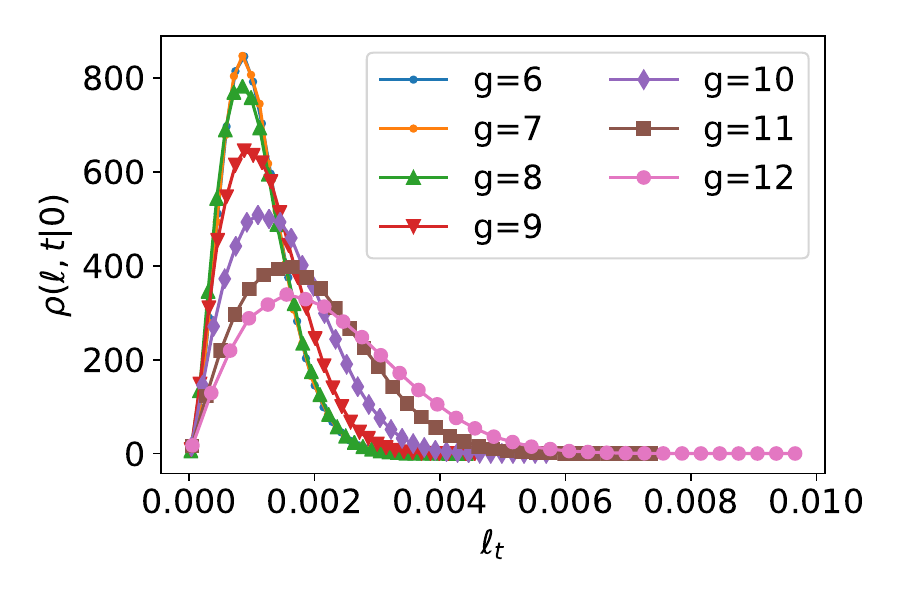} 
\caption{Comparison of empirical PDFs of the boundary local time $\ell_t$ obtained by Monte Carlo simulations in the Koch snowflake with $L=2$, $D=1$, $t = 10^{-7}$, and $N = 10^8$ particles at different generations: $g=6, 7$ (dots), $g=8$ (up triangles), $g=9$ (down triangles), $g=10$ (diamonds), $g=11$ (squares), $g=12$ (circles). 
}
\label{fig:histgfcn}
\end{figure}

At very short times ($t \ll h_g^2/D$), the local environment of the Koch snowflake is identical to the vicinity of the vertex of a wedge of angle $\pi/3$, and the distribution of the boundary local time should be identical to that in the wedge. 
In this regime, the distribution does not depend on $g$, given that the condition $t \ll h_g^2 / D$ is satisfied. 
As soon as $t \gtrsim h_g^2/D$, the particle starts to ``feel'' irregularities of the local environment, and the distribution of $\ell_t$ starts to depend on $g$. 
This is confirmed by Fig. \ref{fig:histgfcn} that presents empirical PDFs of $\ell_t$ at $t = 10^{-7}$ (and $D = 1$). 
For generations $g \le 7$, $h_g^2/D = 4 \times 9^{-g}$ remains much larger than $t = 10^{-7}$, and the PDFs are identical to those of the wedge of angle $\pi/3$. 
In turn, for larger $g$, we observe the effect of the generation on the PDFs: higher generations lead to broader distributions.

\begin{figure}[t!]
\centering
\includegraphics[width=0.7\linewidth]{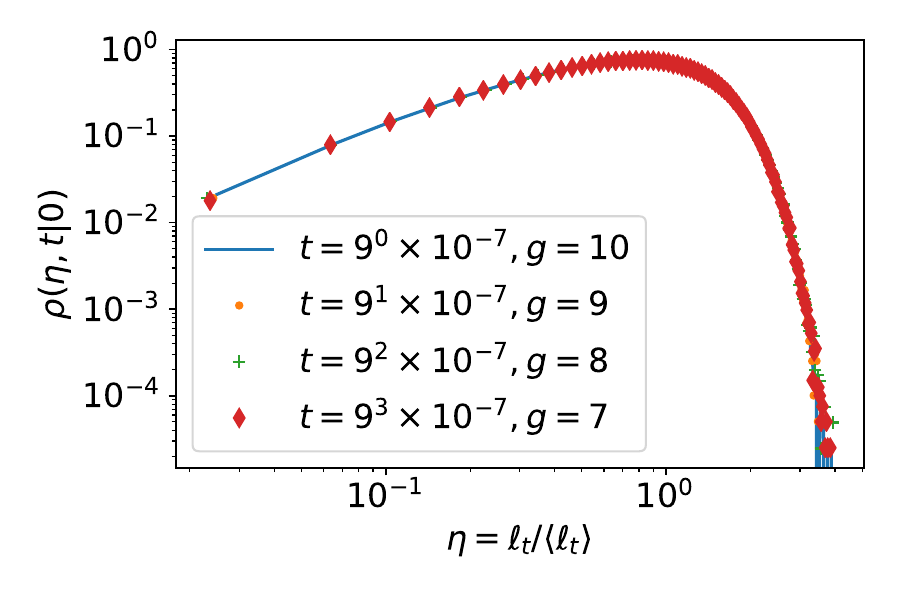} 
\caption{Comparison of empirical PDFs (in log-log scale) of the boundary local time $\ell_t$ normalized by its mean value, obtained in the Koch snowflake with $L=2$, $D=1$, and $N = 10^8$ particles by Monte Carlo simulations at different times and generations: $g = 10$, $t = 10^{-7}$ (blue line), $g = 9$, $t = 9 \times 10^{-7}$ (orange dot), $g = 8$, $t = 9^2 \times 10^{-7}$, and $g=7$, $t = 9^3 \times 10^{-7}$ (red diamond). 
}
\label{fig:histsimilar}
\end{figure}

As the boundary is self-similar, an appropriate rescaling of time $t$ should keep the distribution of the boundary local time invariant. 
In fact, let us consider the distribution of $\ell_t$ for a given generation $\Omega_g$. The next generation $\Omg_{g+1}$ differs from $\Omg_g$ by adding the geometric details at the new smallest scale $h_{g+1} = h_g / 3$. If time $t$ is reduced by a factor of 9, the particle diffusing in $\Omg_{g+1}$ would effectively explore the environment of $\Omg_g$. In other words, the PDF of $\ell_{t/9}$ in $\Omg_{g+1}$ is expected to be identical to the PDF of $\ell_t / 3$ in $\Omg_g$. This self-similarity is indeed observed in Fig. \ref{fig:histsimilar}.  
Note that $\ell_t$ was rescaled by its mean value $\lla \ell_t \rra_0$. 

\begin{figure}[t!]
\centering
\includegraphics[width=0.7\linewidth]{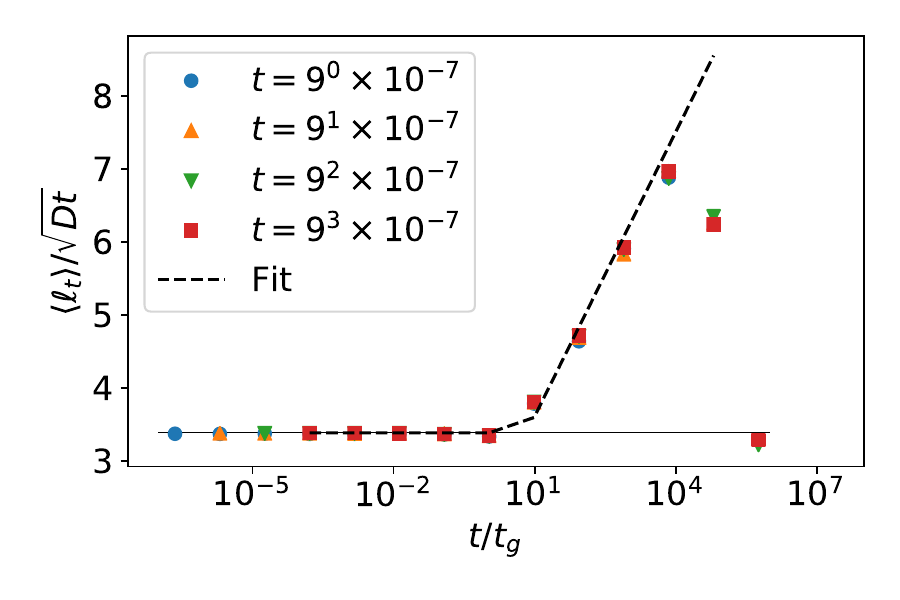} 
\caption{
Mean boundary local time (rescaled by $\sqrt{Dt}$) obtained in the Koch snowflake with $L=2$, $D=1$, and $N = 10^8$ particles by Monte Carlo simulations for different times $t$ and different generations $g$. 
The dashed line presents an empirical fitting. 
}
\label{fig:histnorm}
\end{figure}

Let us now inspect the mean value of the boundary local time. A systematic study was done for the Koch snowflake at different times $t$ and generations $g$. In order to compare these cases, we introduce the time scale $t_g = h_g^2 / D = L^2 / (9^g D)$, and consider $\lla \ell_t \rra_0$ as a function of $t/t_g$ (Fig. \ref{fig:histnorm}). 
For small $g$ and thus small $t/t_g$, the particle explores the local environment of the wedge, yielding the rescaled mean value $\langle \ell_t \rangle_0 / \sqrt{Dt} = 6/\sqrt{\pi} \approx 3.4$, which is shown by a horizontal line.  
When $t$ becomes comparable to $t_g$, the irregularity of the boundary starts to affect the statistics of encounters, yielding an increase of the rescaled mean value. 
This increase is observed up to $t/t_g \approx 10^4$.  
Fitting $\lla \ell_t \rra_0 / \sqrt{Dt} = f(t/t_g)$ in this region, we find an approximation 
\begin{equation}
f(z) = \frac{6}{\sqrt{\pi}} \lls 1 + \frac{\ln(0.15z)}{6}  \Theta(z - 1/0.15) \rrs \,,
\end{equation} 
where $\Theta(x)$ is the Heaviside step function. 
When $t/t_g$ exceeds $10^4$, the rescaled mean value starts to decrease. This behavior qualitatively agrees with what we observed for the simplified case of a wedge with channels: the mean boundary local time first increases and then decreases (Fig. \ref{fig:ellm_aeqb}). 
Remarkably, the effect of the Koch snowflake complexity is rather weak: the ratio $\lla \ell_t \rra_0 / \sqrt{Dt}$ varies by a factor of 2, whereas $t/t_g$ changes over many orders of magnitude. 
A similar behavior was observed for other Koch snowflakes of different angles $\alpha$ (results are not shown). 
As $\alpha$ increases, the maximum is shifted to larger times, e.g., $t/t_g \approx 10^6$ for $\alpha = 3\pi/4$.

\subsection{Local persistence exponents}
\label{sec:LPEk}

\begin{figure}[t!]
\centering
\includegraphics[width=0.7\linewidth]{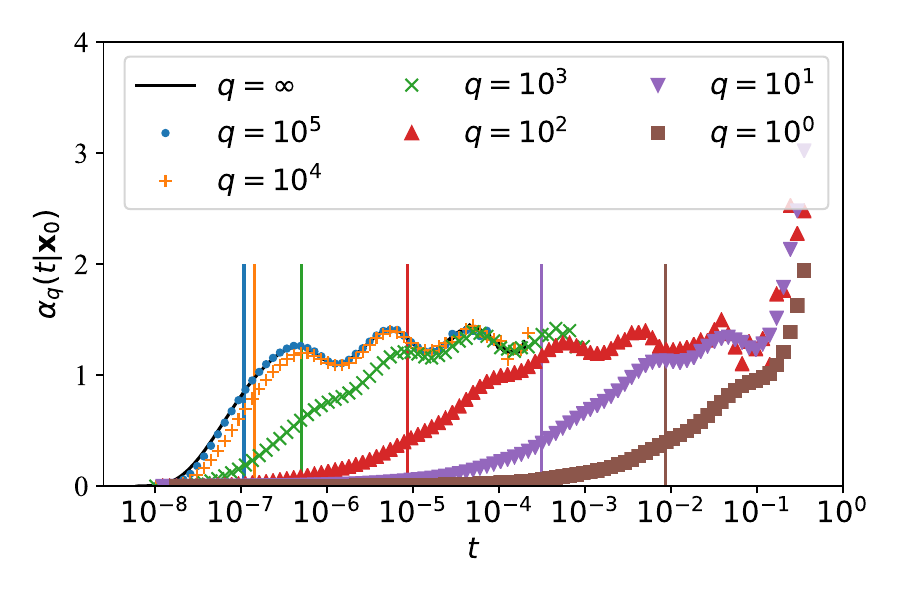} 
\caption{Comparison of local persistence exponents $\alpha_q(t|\x_0)$ obtained in the sixth generation $\Omg_{6}$ of the Koch snowflake with $L=2$, $D=1$, and $N = 10^8$ particles by Monte Carlo simulations for different reactivities: $q=\infty$ (thin black line), $q=10^5$ (blue dots), $q=10^4$ (orange plus), $q=10^3$ (green crosses), $q=10^2$ (red up triangles), $q=10^1$ (violet down triangles), and $q=10^0$ (brown squares). The median time $T_q$ for each reactivity (from $q = 10^5$ to $q = 10^{0}$) is indicated by the vertical line with the corresponding color: $1.09 \times 10^{-7}$, $1.44 \times 10^{-7}$, $5.06 \times 10^{-7}$, $8.66 \times 10^{-6}$, $3.13 \times 10^{-4}$, $8.60 \times 10^{-3}$. }
\label{fig:lpe}
\end{figure}

In many complex systems, the survival probability exhibits a power-law decay (see \cite{levitz2006brownian, Bray13, Levernier19} and references therein). 
In our previous work \cite{ye2024first}, we studied the long-time asymptotic behavior of the survival probability $S_\infty(t|\x_0)$ in the Koch snowflake with a perfectly reactive boundary. For this purpose, the local persistence exponent (LPE) was introduced as the negative logarithmic derivative of the survival probability that we extend here to the case of finite reactivity: 
\begin{equation}
\alpha_q(t|\x_0) = - \frac{\partial \ln S_q(t|\x_0)}{\partial \ln t} \,.
\end{equation} 
Due to the self-similarity of the Koch snowflake boundary, $\alpha_\infty (t|\x_0)$ was shown to exhibit log-periodic oscillations in time. 
Here we inspect how the finite reactivity of the boundary can affect this behavior.

We compute the LPE of the survival probability $S_q(t|\x_0)$ by estimating the latter via the EFS approach. 
The starting point is located at $\x_0 = (x_0, y_0 + 10^{-3})$, where $(x_0,y_0)$ is the bottom vertex of the Koch snowflake as shown in Fig. \ref{sub:Kochlocal2}. 
Here, the starting point is not located at the vertex, in order to enable comparison with $\alpha_\infty(t|\x_0)$ for the perfectly reactive case. 
To avoid too long computational time for small values of $q$, we focus on the generation $g=6$. 
Figure \ref{fig:lpe} compares LPEs for different reactivities $q$ from infinity to $1$.

As discussed in \cite{ye2024first}, $\alpha_\infty(t|\x_0)$ exhibits a transient regime of monotonous growth at very short times, log-periodic oscillations at intermediate times due to self-similarity of the boundary, and further linear increase with $t$ at long times due to confinement in a bounded domain. We observe here similar trends for $\alpha_q(t|\x_0)$. When the reactivity is very high ($q = 10^5$), the particle reacts after first few encounters with the boundary, so that $S_q(t|\x_0)$ is close to $S_\infty(t|\x_0)$, implying $\alpha_q(t|\x_0) \approx \alpha_\infty(t|\x_0)$. 
A decrease of reactivity extends the transient regime to larger and larger times. Moreover, the ultimate linear increase, $\alpha_q(t|\x_0) \propto t$ starts at $t \approx L^2/D$, independently of $q$. As a consequence, the range of log-periodic oscillations is reduced as $q$ decreases. For instance, there is no such an intermediate regime for $q = 1$. 
Note that 
a natural timescale of the transient regime is the median reaction time $T_q$, at which half of the particles survive $S_q(T_q|\x_0) = 1/2$. In fact, a significant decay of the survival probability is expected at $t \gg T_q$.

\section{Discussion and conclusion}
\label{sec:discon}

In this paper, we studied the boundary local time in polygonal domains. 
First, our theoretical framework outlined the statistics of the boundary local time in a wedge, including the mean value, the variance, and the asymptotic behavior of the probability density. 
In fact, Eqs. (\ref{eq:ellnewmean}, \ref{eq:ellnewvar}) give the mean value and the variance of the boundary local time when the starting point is located at the vertex, whereas their extensions to arbitrary starting points are given in \ref{sec:appMBLT}; note that the prefactors $v_\alpha$ in Eq. (\ref{eq:ellnewvar}) were found explicitly for the angles $\alpha = \pi/n$ with an integer $n$, whereas their computation for any $\alpha$ is possible via coupled PDEs between moments $\lla \ell_t^k \rra_{\x_0}$ from \ref{sec:appPDE}. 
The asymptotic behavior (\ref{eq:asymlong}) of the probability density $\rho_\alpha(\ell,t|\x_0)$ in the limit $\ell \to \infty$ was derived for wedges of angle $\alpha \leq \pi$ from the duality between the Steklov problem and the Robin Laplacian problem. 
As the small-$\ell$ behavior relies on all Steklov eigenmodes, we could not establish this behavior rigorously, but we provided its conjectural form.

To deal with polygonal domains, we developed an efficient escape-from-a-sector approach, which relies on the approximate solution of the escape problem for a sector. Comparison of simulated values of $\ell_t$ with theoretical predictions in wedges helped us to validate the EFS approach and to access its accuracy. 
The EFS approach was also verified by computing the spread harmonic measure distribution in the Koch snowflake and comparing with an alternative computation by a FEM. 

The non-monotonous behavior of the mean boundary local time in complex domains presents one of the main numerical results of the paper. 
A compromise between an increase of the perimeter and its reduced accessibility due to larger spaces 
for diffusion was identified as its origin. 
Indeed, the larger perimeter tends to increase the mean boundary local time, while the larger space for diffusion diminishes the chance of encounters on the surface in a given time. 
Such a non-monotonous dependence was first observed for a simple model of the wedge of angle $\pi/2$ with two rectangular channels and then for finite generations of the Koch snowflake. 
In the latter case, a clear increase of the mean boundary local time was noticed for $t/t_g \in (0,10^4)$, which is followed by a decrease. The statistics of encounters is much richer due to the self-similarity of the boundary. 
When the starting point is located at the bottom vertex, changing the generation $g$ is equivalent to rescaling time $t$, yielding identical PDFs of $\ell_t$ after normalization by the mean value $\lla \ell_t \rra_0$.

Self-similarity of the Koch snowflake boundary leads to log-periodic oscillations of the local persistence exponent of the survival probability. We examined how the finite reactivity affects this behavior. 
Three regimes were distinguished: (i) a transient growth  at very short times; (ii) log-oscillations at intermediate times; (iii) linear growth at long times due to confinement in a bounded domain. 
We introduced the median time $T_q$ as a suitable timescale for the transient regime. 
When $t \lesssim T_q$, the number of encounters of the diffusing particle with the boundary is not enough to ensure the reaction event, and the survival probability is close to $1$, yielding $\alpha_q(t|\x_0) \approx 0$. In turn, the asymptotic decay of the survival probability is observed at times $t \gg T_q$. As a consequence, when the reactivity $q$ is small, $T_q$ is large, and the intermediate regime with log-periodic oscillations disappears. In practice, both geometry and reactivity affect the survival probability, its asymptotic behavior, and the median time $T_q$. Their systematic analysis presents an interesting perspective of this work. 

\section*{Acknowledgements}
D.S.G. acknowledges the Simons Foundation for supporting his sabbatical sojourn in 2024 at the CRM (CNRS - University of Montreal, Canada), as well as the Alexander von Humboldt Foundation for support within a Bessel Prize award.
Y.Y acknowledges Adrien Chaigneau and Xiaozhen Wang for fruitful discussions about the asymptotic behavior. 
We also thank Adrien Chaigneau for his computations of the spread harmonic measure on the Koch snowflake by a FEM (Fig. \ref{fig:shm}) for validation purposes.

\appendix

\section{Coupled PDEs for the moments}
\label{sec:appPDE}

The moments of the boundary local time can be found from
the propagator $G_0(\x,t|\x_0)$ via a hierarchical set of PDEs.  Their
derivation relies on the relation (\ref{eq:Sqmg}) between the survival
probability and the probability density function of $\ell_t$. 
The survival probability $S_q(t|\x_0)$ that satisfies Eqs. (\ref{eq:survprob}), admits the following Taylor expansion
\begin{equation}
S_q(t|\x_0) = \sum\limits_{k=0}^\infty \frac{(-q)^k L_k(t|\x_0)}{k!} \,,
\end{equation}
where $L_k(t|\x_0) = \langle \ell_t^k\rangle_{\x_0}$ denotes the $k$-th order
moment of $\ell_t$.  Substituting this Taylor expansion into
Eqs. (\ref{eq:survprob}), we get a sequence of PDEs for the moments with $k
= 1,2,\ldots$
\begin{equation}
\partial_t L_k(t|\x_0) = D\Delta L_k(t|\x_0) \quad \textrm{in}~\Omega, \quad
\partial_n L_k(t|\x_0) = k L_{k-1}(t|\x_0)  \quad \textrm{on}~\pa, \quad L_k(0|\x_0) = 0.
\end{equation}
Note that for $k=0$, $L_0(t|\x_0) = 1$. 
Equivalently, the Laplace transform of $L_k(t|\x_0)$, $\tilde{L}_k(p|\x_0) = \int_0^\infty \md t \  e^{-pt} L_k(t|\x_0)$, satisfies 
\begin{equation}  \label{eq:Ln_tilde}
(p - D\Delta) \tilde{L}_k(p|\x_0) = 0 \quad \textrm{in}~\Omega, \qquad
\partial_n \tilde{L}_k(p|\x_0) = k \tilde{L}_{k-1}(p|\x_0)  \quad \textrm{on}~\pa.
\end{equation}
Let $\tilde{G}_0(\x,p|\x_0)$ denote the Laplace transform of the
propagator $G_0(\x,t|\x_0)$ with the Neumann boundary condition that satisfies
\begin{equation}  \label{eq:G0_tilde}
(p - D\Delta) \tilde{G}_0(\x,p|\x_0) = \delta(\x-\x_0) \quad \textrm{in}~\Omega, \qquad
\partial_n \tilde{G}_0(\x,p|\x_0) = 0  \quad \textrm{on}~\pa.
\end{equation}
Multiplying Eq. (\ref{eq:Ln_tilde}) by $\tilde{G}_0(\x,p|\x_0)$,
multiplying Eq. (\ref{eq:G0_tilde}) by $\tilde{L}_k(p|\x_0)$,
subtracting these equations, integrating them over $\x_0\in\Omega$, and using the Green's formula, we get
\begin{equation}  \label{eq:Ltilde_set}
\tilde{L}_k(p|\x) = k \int\limits_{\pa} \md\x_0 \, D\tilde{G}_0(\x,p|\x_0) \, \tilde{L}_{k-1}(p|\x_0).
\end{equation}
In the time domain, this equation implies:
\begin{equation} \label{eq:appeqA6}
L_k(t|\x) = k \int\limits_{\pa} \md\x_0 \, \int\limits_0^t \md t^\prime \, D G_0(\x,t^\prime | \x_0) L_{k-1} (t - t^\prime | \x_0) \,.
\end{equation}
As $L_0(t|\x_0) = 1$ and thus $\tilde{L}_0(p|\x_0) = 1/p$, we have
\begin{equation}
\tilde{L}_1(p|\x) = \frac{1}{p} \int\limits_{\pa} \md\x \, D\tilde{G}_0(\x,p|\x_0) ,
\end{equation}
and we therefore retrieve the relation (\ref{eq:ellG}): 
\begin{equation}
L_1(t|\x_0) = \int\limits_0^t \md t' \, \int\limits_{\pa} \md\x \, DG_0(\x,t'|\x_0) .
\end{equation}
In turn, the second moment reads
\begin{align}
L_2(t|\x_0) & = 2\int\limits_0^t \md t' \, \int\limits_{\pa} \md\x \, DG_0(\x,t'|\x_0) \, L_1(t-t'|\x) \label{eq:appB10} \\ 
& = 2D^2 \int\limits_{\pa} \md\x_1 \, \int\limits_{\pa} \md\x_2 \, \int\limits_0^t \md t_1 \, \int\limits_0^{t_1} \md t_2 \,
G_0(\x_2,t_2|\x_1) \, G_0(\x_1,t-t_1|\x_0). \label{eq:appB11}
\end{align}

If the boundary $\pa$ of the domain is bounded, one can use the
spectral decomposition of $\tilde{G}_0(\x,p|\x_0)$ over the Steklov
eigenfunctions $V_k^{(p)}$, which reads for $\x\in\pa$ as \cite{Grebenkov20}
\begin{equation}
D\tilde{G}_0(\x,p|\x_0) = \sum\limits_k \frac{V_k^{(p)}(\x_0) V_k^{(p)}(\x)}{\mu_k^{(p)}} \,.
\end{equation}
The orthogonality of Steklov eigenfunctions on the boundary allows one
to compute integrals over $\pa$ in Eq. (\ref{eq:Ltilde_set}) and thus
to express the Laplace transform of the $k$-th moment as
\begin{equation}
\tilde{L}_k(p|\x_0) = \frac{1}{p} \sum\limits_k \frac{V_k^{(p)}(\x_0)}{[\mu_k^{(p)}]^k} \int\limits_{\pa} \md\x \, V_k^{(p)}(\x)\,,
\end{equation}
i.e., we retrieved the spectral expansion from \cite{Grebenkov19}.
However, this computation is not applicable when the boundary $\pa$ is unbounded (as in the case of wedges), and one needs to evaluate the integrals in Eq. \ref{eq:appeqA6}.

\section{Explicit results for a quadrant}
\label{app:quadrant}

For completeness, we provide some explicit results for the quadrant
$\R_{+}^2$, i.e., the wedge of angle $\pi/2$.  The following computation is elementary and relies on the independence of horizontal
and vertical displacements that reduces the analysis to
one-dimensional problems. 
Despite their simplicity, we are not aware of earlier references presenting these results. 

We recall that the probability density of the boundary local time
$\ell_t$ on the positive semi-axis reads as
\begin{equation}
\rho_{\pi}(\ell,t|x_0) = \erf\left(\frac{x_0}{\sqrt{4Dt}}\right) \delta(\ell) + \frac{e^{-(\ell+x_0)^2/(4Dt)}}{\sqrt{\pi Dt}}  \,,
\end{equation}
where the first {\it singular} term, $\delta(\ell)$, accounts for
trajectories that do not hit the boundary up to time $t$, with
$\erf(x_0/\sqrt{4Dt})$ being the associated survival probability. As
a consequence, we have
\begin{align*}
\rho_{\pi/2}(\ell,t|x_0,y_0) & = \int\limits_0^\infty \md\ell_1 \, \rho_{\pi}(\ell_1,t|x_0) 
\int\limits_0^\infty \md\ell_2 \, \rho_{\pi}(\ell_2,t|y_0) \, \delta(\ell_1+\ell_2-\ell) \\
& = S_x S_y \delta(\ell) + \frac{S_x}{\sqrt{\pi Dt}} e^{-(\ell+y_0)^2/(4Dt)}
+ \frac{S_y}{\sqrt{\pi Dt}} e^{-(\ell+x_0)^2/(4Dt)} \\
& \hskip 4mm  + \frac{2}{\pi \sqrt{Dt}} \int\limits_0^\infty \md z_1 \, \int\limits_0^\infty \md z_2 \,
e^{-(z_1+\xi_1)^2-(z_2+\xi_2)^2} \delta(z_1+z_2-\ell/\sqrt{4Dt}),
\end{align*} 
where $\xi_1 = x_0/\sqrt{4Dt}$, $\xi_2 = y_0/\sqrt{4Dt}$, $S_x =
\erf(x_0/\sqrt{4Dt})$, and $S_y = \erf(y_0/\sqrt{4Dt})$.  Evaluating
the last integral, we have 
\begin{align} \nonumber
& \rho_{\pi/2}(\ell,t|x_0,y_0) =  \erf\llp\frac{x_0}{\sqrt{4D t}}\rrp \erf\llp\frac{y_0}{\sqrt{4D t}}\rrp \delta(\ell) \\
\nonumber & + \frac{e^{-(\ell+y_0)^2/(4Dt)}}{\sqrt{\pi Dt}} \erf\left(\frac{x_0}{\sqrt{4Dt}}\right) + \frac{e^{-(\ell+x_0)^2/(4Dt)}}{\sqrt{\pi Dt}} \erf\left(\frac{y_0}{\sqrt{4Dt}}\right) \\
& + \frac{e^{-(\ell+x_0+y_0)^2/(8Dt)}}{\sqrt{2\pi Dt}} 
\biggl[\erf\left( \frac{\ell-y_0+x_0}{\sqrt{8Dt}}\right) + \erf\left( \frac{\ell+y_0-x_0}{\sqrt{8Dt}}\right)\biggr].
\end{align}
%
%
%
%
As the boundary local times on the horizontal and vertical axes are
independent, we get
\begin{align}
\langle \ell_t \rangle_{\x_0} & 
 = \frac{2\sqrt{Dt}}{\sqrt{\pi}} \biggl[e^{-x_0^2/(4Dt)} -  \frac{\sqrt{\pi} \,x_0}{\sqrt{4Dt}}\, \erfc\left(\frac{x_0}{\sqrt{4Dt}}\right)
+ e^{-y_0^2/(4Dt)} - \frac{\sqrt{\pi} \,y_0}{\sqrt{4Dt}}\, \erfc\left(\frac{y_0}{\sqrt{4Dt}}\right)\biggr].
\end{align}
This relation agrees with our general expression
(\ref{eq:L1_special}), with $m = 2$, in which two terms in the sum
can be associated to $y_0 = r_0 \sin\theta_0$ and $x_0 = r_0 \sin
\theta_1 = r_0 \cos \theta_0$.


We also note that the probability density $\rho_\alpha(\ell,t|\x_0)$ of the
boundary local time $\ell_t$ is tightly related to the probability
density $U_\alpha(\ell,t|\x_0)$ of the first-crossing time $\T_\ell = \inf\{
t > 0 ~:~ \ell_t > \ell\}$ of a threshold $\ell$ \cite{Grebenkov20}: 
\begin{equation}
\int\limits_\ell^\infty \md\ell' \, \rho_\alpha(\ell',t|\x_0) = \P_{\x_0}\{\ell_t > \ell\} 
= \P_{\x_0}\{ \T_\ell < t\} = \int\limits_0^t \md t' \, U_\alpha(\ell,t'|\x_0),
\end{equation}
i.e.,
\begin{equation}
U_\alpha(\ell,t|\x_0) = \int\limits_\ell^\infty \md\ell' \, \partial_t \rho_\alpha(\ell',t|\x_0).
\end{equation}
For instance, for the wedge of angle $\pi$, one has
\begin{equation}
U_\pi(\ell,t|0) = \frac{\ell \, e^{-\ell^2/(4Dt)}}{\sqrt{4\pi Dt^3}} \,.
\end{equation}
Similarly, we get for the quadrant
\begin{equation}
U_{\pi/2}(\ell,t|0) = \frac{\ell \, e^{-\ell^2/(8Dt)} \erf(\ell/\sqrt{8Dt})}{\sqrt{2\pi Dt^3}} \,.
\end{equation}
In both cases, the mean first-crossing time is infinite, as expected
from the divergence of the mean first-passage time.

\section{Mean boundary local time and its variance in wedges}
\label{sec:appMBLT}

In this Appendix, we elaborate on the computation of the mean boundary local time (\ref{sec:appA2}) and its variance (\ref{sec:appA3}). 
These computations require knowledge of the propagator for the wedge.  Even though the computation of the propagator is standard, we reproduce it in \ref{sec:app_prop} for completeness.

\subsection{Propagator in the wedge}
\label{sec:app_prop}
Let us compute the propagator $G_0(\x,t|\x_0)$ in a wedge of angle
$\alpha$ with Neumann boundary condition.  Its Laplace transform, $\tilde{G}_0(\x,p|\x_0) = \int\nolimits_0^\infty \md t \, e^{-pt} G_0(\x,t|\x_0)$, satisfies
\begin{align} \nonumber
(p - D\Delta) \tilde{G}_0(\x,p|\x_0) &= \delta(\x-\x_0) = r_0^{-1} \delta(r-r_0) \delta(\theta-\theta_0) \quad \textrm{in}~\Omega, \\
\qquad \partial_n \tilde{G}_0(\x,p|\x_0) &= 0 \quad \textrm{on}~\pa.
\end{align}
One can search for its solution in the form that respects the boundary
condition:
\begin{equation}
\tilde{G}_0(\x,p|\x_0) = \frac{1}{\alpha} \tilde{g}_0(r,p|r_0) + \frac{2}{\alpha}
\sum\limits_{n=1}^\infty \cos(\nu_n\theta) \cos(\nu_n\theta_0) \tilde{g}_n(r,p|r_0),
\end{equation}
with $\nu_n = \pi n/\alpha$ and unknown radial functions
$\tilde{g}_n(r,p|r_0)$.  Substitution of this form into the above
equation yields
\begin{equation}
\L_0 \tilde{g}_0(r,p|r_0) + 2\sum\limits_{n=1}^\infty \cos(\nu_n\theta) \cos(\nu_n\theta_0) \L_n \tilde{g}_n(r,p|r_0) 
= \frac{\alpha}{D r_0} \delta(r-r_0) \delta(\theta-\theta_0),
\end{equation}
where $\L_n = p/D - (\partial_r^2 + r^{-1} \partial_r - \nu_n^2
r^{-2})$.  This equation can be satisfied by imposing
\begin{equation}
\L_n \tilde{g}_n(r,p|r_0) = \frac{1}{D r_0} \delta(r-r_0).
\end{equation}
For each $n$, the solution of this equation can be found by solving
the homogeneous equation for $r < r_0$ and $r > r_0$ and then matching
two solutions.  One gets
\begin{equation}
\tilde{g}_n(r,p|r_0) = \frac{1}{D}  I_{\nu_n}(r_< \sqrt{p/D}) K_{\nu_n}(r_> \sqrt{p/D}) ,
\end{equation}
where $r_< = \min\{r,r_0\}$, $r_> = \max\{r,r_0\}$, and $I_{\nu_n}(z)$ and
$K_{\nu_n}(z)$ are the modified Bessel functions of the first and second
kind.  The inverse Laplace transform of this function yields
\cite{Carslaw}
\begin{equation}
g_{n}(r,t|r_0) = \frac{1}{2Dt} e^{-(r^2+r_0^2)/(4Dt)} I_{\nu_n}(r r_0/(2Dt)) .
\label{eq:appa6}
\end{equation}
We conclude that
\begin{equation}
G_0(\x,t|\x_0) = \frac{g_0(r,t|r_0)}{\alpha}  + \frac{2}{\alpha}\sum\limits_{n=1}^\infty \cos(\nu_n\theta)\cos(\nu_n\theta_0) g_{n}(r,t|r_0).
\label{eq:appa7}
\end{equation}

\subsection{Mean boundary local time}
\label{sec:appA2}
The substitution of Eq. (\ref{eq:appa7}) into Eq. (\ref{eq:ellG}) yields the mean boundary local time 
%
\begin{align}
\nonumber \langle \ell_t\rangle_{\x_0} &= D\int\limits_0^t \md t' \, \int\limits_0^\infty \md r \, \biggl[G_0(\x,t'|\x_0)|_{\theta = 0} + 
G_0(\x,t'|\x_0)|_{\theta = \alpha}\biggr] \\
& = \frac{D}{\alpha}\sum\limits_{n=0}^\infty \epsilon_n \cos(\nu_n\theta_0) (1 + (-1)^n) \int\limits_0^t \md t' \,e^{-r_0^2/(8Dt')} 
\frac{\sqrt{\pi} }{2\sqrt{4Dt'}} I_{\nu_n/2}(r_0^2/(8Dt')) ,
\end{align}
with $\epsilon_n = 2-\delta_{n,0}$, and we used
\begin{equation}
\int\limits_0^\infty \md z \, e^{-z^2} \, I_{\nu_n}(az) = \frac{\sqrt{\pi}}{2} e^{a^2/8} I_{\nu_n/2}(a^2/8).
\end{equation}

At $r_0 = 0$, only the term with $n = 0$ contributes, yielding
\begin{equation}
\frac{1}{\alpha} \int\limits_0^{\infty} \md r \, 2g_0(r,t|r_0) = \frac{\sqrt{\pi}}{\alpha \sqrt{Dt}} \,,
\end{equation}
from which the integral over $t$ implies
\begin{equation}
\langle \ell_t \rangle_0 = \frac{\pi}{\alpha} \, \frac{2\sqrt{Dt}}{\sqrt{\pi}} .
\label{eq:A11}
\end{equation} 
We therefore retrieved the expression obtained in the main text by
probabilistic and symmetry arguments.

In turn, for $r_0 > 0$, one can rewrite the above expression as
\begin{align}
\langle \ell_t\rangle_{\x_0} & = \frac{r_0 \sqrt{\pi}}{2\sqrt{2} \alpha}
\sum\limits_{n=0}^\infty \epsilon_n \cos(2\nu_n\theta_0) \int\limits_{r_0^2/(8Dt)}^\infty \frac{\md z}{z^{3/2}}\,e^{-z} I_{\nu_n}(z) .
\label{eq:A12}
\end{align}
Note that an accurate numerical computation of this sum may require
taking a significant number of terms.

\subsubsection*{Special case.}
In the special case $\alpha = \pi/m$ with an integer $m = 1,2,\ldots$,
one can further simplify Eq. (\ref{eq:A12}) to get a closed-form expression.
For this purpose, we use the following representation of the modified
Bessel function of the first kind:
\begin{equation}  \label{eq:BesselI}
I_{\nu_n}(z) = \frac{1}{\pi} \int\limits_0^{\pi} \md\theta \, \cos (\nu_n\theta) e^{z\cos\theta} 
- \frac{\sin(\nu_n \pi)}{\pi} \int\limits_0^{\infty} \md x \, e^{-z\cosh x-\nu_n x} .
\end{equation}
Since $\nu_n = \pi n/\alpha = mn$ is integer, the second term
vanishes, whereas the substitution of the first term into Eq. (\ref{eq:A12})
yields
\begin{equation}
\langle \ell_t\rangle_{\x_0} = \frac{2\sqrt{Dt}}{\sqrt{\pi}} L^{(1)}(\eta) ,
\end{equation}
where $\eta = r_0^2/(8Dt)$, and 
%
%
\begin{equation}
L^{(1)}(\eta) = \frac{\sqrt{\eta}}{2\alpha}\int\limits_0^\pi \md\theta \,\int\limits_\eta^\infty \frac{\md z}{z^{3/2}} \, e^{-z} e^{z\cos\theta} \sum\limits_{n=0}^\infty \epsilon_n \cos(2\nu_n \theta_0) \cos(\nu_n \theta) .
\end{equation}
Using the identity
\begin{equation}
\int\limits_x^\infty \frac{\md z}{z^{3/2}} \, e^{-z} =  \frac{2 e^{-x}}{\sqrt{x}}  - \sqrt{\pi} \, \erfc(\sqrt{x}) \,,
\end{equation}
we get
\begin{align} \nonumber 
L^{(1)}(\eta) = \frac{1}{\alpha}\int\limits_0^\pi \md\theta \, \biggl[e^{-\eta(1-\cos\theta)} 
- \sqrt{\pi} \sqrt{\eta(1-\cos\theta)}\, \erfc(\sqrt{\eta(1-\cos\theta)})\biggr] \\
\times \sum\limits_{n=0}^\infty \epsilon_n \cos(2\pi n \theta_0/\alpha) \cos(\pi n \theta/\alpha).
\label{eq:appA17}
\end{align}
Due to the reflection symmetry with respect to the central ray of the
wedge, we can assume that $2\theta_0 \leq \alpha$.  Setting $\xi_0 =
2\theta_0/\alpha$, we see that $0\leq \xi_0\leq 1$.  In turn, $\xi =
\theta/\alpha$ can take any value from $0$ to $m$ given that $\theta$
varies up to $\pi$ in Eq. (\ref{eq:appA17}).  We aim at applying the following identity
\begin{equation}  \label{eq:cosine_delta}
\sum\limits_{n=0}^\infty \epsilon_n \cos(\pi n \xi_0) \cos(\pi n \xi) = \delta(\xi_0-\xi) \qquad (0 < \xi_0, \xi < 1).
\end{equation}
For this purpose, the integral over $\theta$ from $0$ to $\pi$ can be
split into $m$ integrals over the intervals $(\alpha j,\alpha(j+1))$,
with $j=0,1,\ldots,m-1$.  To apply the above identity, we introduce the angles
\begin{equation}  \label{eq:thetaj}
\theta_j = \begin{cases} \theta_0 + \tfrac12 \alpha j  \hskip 13.3mm \textrm{if $j$ is even},\cr
\tfrac12 \alpha (j+1) - \theta_0 \quad \textrm{if $j$ is odd} , \end{cases}
\end{equation}
for $j = 0, 1,2,\ldots,m-1$.  These angles determine the mirror
reflections of the starting point $(r_0,\theta_0)$ with respect to the
reflecting boundary of the wedge.  Applying
Eq. (\ref{eq:cosine_delta}) with a suitable shift of $\xi$ to ensure
that $\xi - \xi_j \in (0,1)$ for each interval $(\alpha
j,\alpha(j+1))$, we get
\begin{align} \label{eq:L1_m}
L^{(1)}(\eta) & = \sum\limits_{j=0}^{m-1} \biggl[e^{-\eta(1-\cos(2\theta_j))} 
- \sqrt{\pi} \sqrt{\eta(1-\cos(2\theta_j))}\, \erfc\biggl(\sqrt{\eta(1-\cos(2\theta_j))}\biggr)\biggr],
\end{align}
from which
\begin{equation}  \label{eq:L1_special}
\langle \ell_t\rangle_{\x_0} = \frac{2\sqrt{Dt}}{\sqrt{\pi}} \sum\limits_{j=0}^{m-1} \biggl[e^{-r_0^2\sin^2 \theta_j/(4Dt)} 
- \sqrt{\pi} \, \frac{r_0\sin \theta_j}{\sqrt{4Dt}}\, \erfc\left(\frac{r_0\sin \theta_j}{\sqrt{4Dt}}\right)\biggr] .
\end{equation}
In the limit $r_0\to 0$, the sum is simply equal to $m = \pi/\alpha$,
and we retrieve Eq. (\ref{eq:A11}).

\subsubsection*{General case.}

Let us first focus on the case $\alpha > \pi$.  As previously, we
assume that $\theta_0 \leq \alpha/2$ (otherwise $\theta_0$ can be
replaced by $\alpha - \theta_0$ by symmetry).  As $\pi/\alpha$ is not
an integer anymore, both integrals in Eq. (\ref{eq:BesselI}) have to be
considered.  The contribution of the first integral is still given by
Eq. (\ref{eq:L1_m}) with $m = 1$ so that only one term is present.
However, the contribution of this single term is canceled if
$2\theta_0 > \pi$, due to the Dirac distribution in
Eq. (\ref{eq:cosine_delta}) and the integration over $\theta$ from $0$ to
$\pi$.  We conclude that
\begin{equation}    \label{eq:L1_auxil1}
L^{(1)}(\eta) = \Theta(\pi - 2\theta_0) \biggl[e^{-2\eta \sin^2 \theta_0} 
- \sqrt{2\pi \eta} \sin \theta_0\, \erfc\bigl(\sqrt{2\eta} \sin\theta_0\bigr)\biggr].
\end{equation}

For the second term, we have
\begin{align*}
L^{(2)}(\eta) & = \frac{\sqrt{\eta}}{2\alpha} \int\limits_0^\infty \md x \, \int\limits_\eta^\infty \frac{\md z}{z^{3/2}} \, e^{-z} e^{-z\cosh x}
\sum\limits_{n=1}^\infty 2 \cos(2\nu_n \theta_0) \sin(\nu_n \pi) e^{-\nu_n x}.
\end{align*}
Using the identity
\begin{equation}
\sum\limits_{n=1}^\infty \sin(n\beta_\pm) e^{-n\delta} 
= \frac{e^{-\delta} \sin \beta_\pm}{1 - 2\cos\beta_\pm e^{-\delta} + e^{-2\delta}} \,,
\end{equation}
with $\delta = \pi x/\alpha$ and $\beta_\pm = \pi(\pi \pm
2\theta_0)/\alpha$, we find
\begin{align*}
L^{(2)}(\eta) = & \frac{1}{\alpha}\int\limits_0^\infty \md x \, \biggl[\frac{e^{-\pi x/\alpha} \sin \beta_+}{1 - 2\cos\beta_+ e^{-\pi x/\alpha} 
+ e^{-2\pi x/\alpha}}
+ \frac{e^{-\pi x/\alpha} \sin \beta_{-}}{1 - 2\cos\beta_{-} e^{-\pi x/\alpha} + e^{-2\pi x/\alpha}}\biggr]  \\
& \times \biggl[e^{-\eta(1+\cosh x)} - \sqrt{\pi \eta(1+\cosh x)} \, \erfc(\sqrt{\eta(1+\cosh x)})\biggr].
\end{align*}
Changing the integration variable, $z = e^{-\pi x/\alpha}$, we 
have
\begin{align*}
L^{(2)}(\eta) = & \frac{1}{\pi}\int\limits_0^1 \md z \, \biggl(\frac{\sin \beta_+}{1 - 2z\cos\beta_+ + z^2}
+ \frac{\sin \beta_{-}}{1 - 2z\cos\beta_{-} + z^2}\biggr)  \\
& \times \biggl[e^{-(z^{\gamma} + z^{-\gamma})^2 \eta/2} - \sqrt{\pi \eta/2} (z^{\gamma} + z^{-\gamma}) \, 
\erfc\bigl((z^{\gamma} + z^{-\gamma})\sqrt{\eta/2}\bigr)\biggr],
\end{align*}
where $\gamma = \alpha/(2\pi)$.  Combining this expression with
Eq. (\ref{eq:L1_auxil1}), we get a relatively simple expression for an
accurate computation of the mean boundary local time:
\begin{equation}  \label{eq:ellm_general}
\langle \ell_t \rangle_{\x_0} = \frac{2\sqrt{Dt}}{\sqrt{\pi}} \bigl(L^{(1)}(\eta) - L^{(2)}(\eta)\bigr).
\end{equation}

Finally, when $\alpha < \pi$ but the ratio $\pi/\alpha$ is not
integer, the second contribution $L^{(2)}(\eta)$ remains unchanged.
In turn, the first contribution admits a form similar to
Eq. (\ref{eq:L1_m}), in which $m = 1 + \lfloor
\pi/\alpha^{+0}\rfloor$, where $\lfloor x\rfloor$ is the integer part
of $x$ (i.e., the greatest integer less than or equal to $x$); here
$\alpha^{+0}$ is the convention to take the limit of $\lfloor
\pi/(\alpha +\epsilon)\rfloor$ as $\epsilon\to 0$ to naturally
incorporate the case when $\alpha = \pi/m$ with an integer $m$.  The
only difference with Eq. (\ref{eq:L1_m}) is that if $2\theta_j$
exceeds $\pi$, the contribution of the corresponding term should be
canceled, as in Eq. (\ref{eq:L1_auxil1}).  We conclude that
\begin{equation}  \label{eq:L1_m2}
L^{(1)}(\eta) = \sum\limits_{j=0}^{m-1} \Theta(\pi/2 - \theta_j) \biggl[e^{-2\eta \sin^2 \theta_j} 
- \sqrt{2\pi\eta} \sin \theta_j\, \erfc\biggl(\sqrt{2\eta} \sin\theta_j\biggr)\biggr],
\end{equation}
and Eq. (\ref{eq:ellm_general}) is now valid for any $0 < \alpha \leq
2\pi$.  We recall that $\theta_0$ should be understood as
$\min\{\theta_0, \alpha-\theta_0\}$.

\subsection{Computation of the variance for $\alpha = \pi/m$} 
\label{sec:appA3}
Since the first moment $L_1(t|\x_0) = \langle \ell_t\rangle_{\x_0}$ is
known via Eq. (\ref{eq:A12}), we can also compute the second moment by
evaluating the integrals in Eq. (\ref{eq:appB10}).  
However, this computation in the general case is tedious.  For this
reason, we focus on the starting point $\x_0 = 0$ and consider only
the special case $\alpha = \pi/m$ with an integer $m$, for which the
first moment admits a simple form (\ref{eq:L1_special}).  In this
case, Eq. (\ref{eq:appB10}) reads
\begin{align}
L_2(t|0) 
& = 2D \int\limits_0^t \md t' \, \int\limits_0^\infty \md r \, \biggl[G_0(r,0,t'|0) L_1(t-t'|r,0) + G_0(r,\alpha,t'|0) L_1(t-t'|r,\alpha)\biggr] ,
\end{align}
where two terms represent the integrals over two rays at $\theta = 0$
and $\theta = \alpha$, written in polar coordinates.  At the starting
point $\x_0 = 0$, the propagator has a simple form, given by the first
term in Eq. (\ref{eq:appa7}).  As both the propagator and the first moment are
symmetric with respect to the central ray of the wedge, the two terms in Eq. (\ref{eq:A12}) 
are identical. Substituting Eq. (\ref{eq:appa6}) and Eq. (\ref{eq:L1_special}),
we get thus
\begin{align}
\nonumber L_2(t|0)  =& 4D \int\limits_0^t \md t' \, \int\limits_0^\infty \md r \, \frac{e^{-r^2/(4Dt')}}{2Dt' \alpha} 
\biggl\{\frac{2\sqrt{D(t-t')}}{\sqrt{\pi}} \sum\limits_{j=0}^{m-1} \biggl[e^{-r^2\sin^2 \theta_j/(4D(t-t'))}  \\
& - \frac{\sqrt{\pi} r\sin\theta_j}{\sqrt{4D(t-t')}}\, \erfc\left(\frac{r\sin\theta_j}{\sqrt{4D(t-t')}}\right)\biggr]\biggr\} ,
\end{align}
where $\theta_j$ were defined in Eq. (\ref{eq:thetaj}) with $\theta_0
= 0$.  Changing the integration variables $z = r/\sqrt{4Dt'}$ and $\tau = t'/t$, we
have
\begin{align} \nonumber
L_2(t|0) = &\frac{8Dt}{\sqrt{\pi} \alpha} \int\limits_0^1 \frac{\md \tau}{\sqrt{\tau}} \, \sqrt{1-\tau} \int\limits_0^\infty \md z \, e^{-z^2} 
 \sum\limits_{j=0}^{m-1} \biggl[e^{-z^2\sin^2 \theta_j \, \tau/(1-\tau)} \\
& - z \sin \theta_j  \sqrt{\frac{ \pi \tau}{1-\tau}} \, \erfc\left(z\sin \theta_j \sqrt{\frac{\tau}{1-\tau}}\right)\biggr].
\end{align}
Evaluation of the integrals over $z$ yields
\begin{align}
L_2(t|0) & = \frac{4Dt}{\alpha}  \sum\limits_{j=0}^{m-1} 
\int\limits_0^1 \frac{\md \tau}{\sqrt{\tau}} \, \sqrt{1-\tau} \biggl\{ \sqrt{\frac{1-\tau}{1 - \tau\cos^2\theta_j}}
- \sin \theta_j \sqrt{\frac{\tau}{1-\tau}} \biggl(1 - \frac{A_\tau}{\sqrt{1+A_\tau^2}}\biggr)\biggr\} ,
\end{align}
where $A_\tau = \sin \theta_j \sqrt{\tau/(1-\tau)}$.  We get then
\begin{align}
\nonumber L_2(t|0) & = \frac{4Dt}{\alpha}  \sum\limits_{j=0}^{m-1} 
\int\limits_0^1 \frac{\md \tau}{\sqrt{\tau}} \, \biggl\{ \frac{1-\tau}{\sqrt{1 - \tau\cos^2\theta_j}}
- \sin \theta_j \sqrt{\tau} \biggl(1 - \frac{\sqrt{\tau} \sin\theta_j}{\sqrt{1-\tau \cos^2\theta_j}}\biggr)\biggr\} \\
& = \frac{4Dt}{\alpha}  \sum\limits_{j=0}^{m-1} \biggl\{ - \sin \theta_j
+ \int\limits_0^1 \frac{d\tau}{\sqrt{\tau}} \, \sqrt{1 - \tau\cos^2\theta_j} \biggr\} 
 = \frac{4Dt}{\alpha}  \sum\limits_{j=0}^{m-1} \frac{\pi/2 - \theta_j}{\cos\theta_j} 
\end{align}
(note that if $\theta_j = \pi/2$, the ratio is set to $1$).  Recalling
Eq. (\ref{eq:thetaj}), we obtain the explicit form of the second
moment 
\begin{equation}
L_2(t|0) = 4Dt \cdot \sigma_m \,,
\end{equation}
with the prefactor $\sigma_m$ that depends on the parity of $m$:  
\begin{align}
\sigma_m & = m \biggl[\frac{1}{2} + \frac{1}{\pi} + \sum\limits_{j=1}^{(m-2)/2} \frac{1 - 2j/m}{\cos (\pi j/m)}\biggr] 
\quad \textrm{if $m$ is even,} \label{eq:appvar27} \\
\sigma_m & = m \biggl[\frac{1}{2} + \sum\limits_{j=1}^{(m-1)/2} \frac{1 - 2j/m}{\cos (\pi j/m)}\biggr] 
\hskip 14mm \textrm{if $m$ is odd.} 
\label{eq:appvar28}
\end{align}
As a consequence, we get the variance, $\Var_0\{\ell_t\} = L_2(t|0) - \lls L_1(t|0) \rrs^2 = 4Dt(\sigma_m - m^2 / \pi)$, and thus the prefactor $v_\alpha$ from Eq. (\ref{eq:varianceva}) reads  
\begin{equation}  \label{eq:valpha} 
v_\alpha = 2 \, \frac{\sigma_m - m^2/\pi}{m(1 - 2/\pi)} \,.
\end{equation}

\section{Mathematical basis for the EFS approach}
\label{sec:derivation}

In this Appendix, we derive the main formulas needed for the
implementation of the EFS approach. 
We consider a sector of angle $\alpha$ and radius $\ve$, which is defined in polar coordinates as: $\Omega = \{ (r,\tta) : 0 < r < \ve, 0 < \tta < \alpha \}$ (see Fig. \ref{fig:scheme2}). 

\subsection{Mean escape time}

We first look at the mean first-passage time (MFPT) to
the absorbing arc $\pa_D$ of the sector, when the particle started from a point $\x_0 \in \Omega$: $T(\x_0) = \E_{\x_0}\{\tau\}$. This
function satisfies:
\begin{equation}
\Delta T = - \frac{1}{D}, \qquad T_{r=\ve} =0 ,\qquad (\partial_\theta T)_{\theta \in \{0,\alpha\}} = 0,
\end{equation}
which admits an explicit exact solution:
\begin{equation}
\E_{\x_0}\{\tau\} = T = \frac{\ve^2 - r_0^2}{4D} \,.
\end{equation}
Indeed, as the segments $\Gamma$ are considered as reflecting here,
they have no effect onto radial displacements so that one retrieves
the MFPT to the boundary of a disk.

\subsection{Mean escape position}

Next, we need to find the mean escape position on the arc (i.e., the
angle $\theta_\tau$). For this purpose, one can compute the Green's
function satisfying mixed Dirichlet/Neumann boundary conditions:
\begin{equation}
-\Delta G(\x,\x_0) = \delta(\x-\x_0), \qquad G|_{\pa_D} = 0,  \qquad   \partial_n G|_{\Gamma} = 0.
\end{equation}
As this computation is standard, we just sketch the main steps.
We search the Green's function in the form
\begin{equation}
G(\x,\x_0) = g_0(r,r_0) + \sum\limits_{n=1}^\infty \cos(\nu_n \theta) \cos(\nu_n \theta_0) g_n(r,r_0),
\end{equation}
where $\nu_n = \pi n/\alpha$, and $g_n(r,r_0)$ are unknown radial
functions that need to be determined from the equation
\begin{equation}
(\partial_r^2 + r^{-1} \partial_r - \nu_n^2 r^{-2}) g_n(r,r_0) = - \frac{2}{\alpha r_0} \delta(r-r_0).
\end{equation}
For $n > 0$, we search for a solution in a standard way:
\begin{equation}
g_n(r,r_0) = \begin{cases}  A_n r^{\nu_n}  \hskip 34.5mm (r < r_0), \cr C_n( (r/\ve)^{\nu_n} - (r/\ve)^{-\nu_n}) \quad (r > r_0), \end{cases}
\end{equation}
where $A_n$ and $C_n$ are unknown coefficients. This form ensures the
regular behavior at $r = 0$ and vanishing of $G$ at $r = \ve$. Matching
the two parts at $r = r_0$ (by requiring the continuity of $G$ and the
drop of its derivative), we get
\begin{equation}
C_n = \frac{a_n r_0}{2\nu_n} (r_0/\ve)^{\nu_n},  \qquad A_n = \frac{C_n}{r_0^{\nu_n}} \bigl[(r_0/\ve)^{\nu_n} - (r_0/\ve)^{-\nu_n}\bigr] ,
\end{equation}
for any $n > 0$. In turn, for $n = 0$, we search $g_0(r,r_0)$ as
\begin{equation}
g_0(r,r_0) = \begin{cases} A_0   \hskip 17.4mm (r < r_0), \cr  C_0 \ln(r/\ve) \quad (r > r_0). \end{cases}
\end{equation}
After matching these solutions at $r = r_0$, we get
\begin{equation}
g_0(r,r_0) = \frac{1}{\alpha} \times \begin{cases} \ln(\ve/r_0)   \quad (r < r_0), \cr  \ln (\ve/r)  \hskip 5.7mm (r > r_0). \end{cases}
\end{equation}
As a consequence, we have
\begin{equation}  \label{eq:Green_DN}
G(\x,\x_0) = g_0(r,r_0) + \sum\limits_{n=1}^\infty \frac{1}{\pi n} \cos(\nu_n \theta) \cos(\nu_n \theta_0)
\times \begin{cases}  \bigl[(r/r_0)^{\nu_n} - (r r_0/\ve^2)^{\nu_n}\bigr]  \quad (r < r_0), \cr
 \bigl[(r_0/r)^{\nu_n} - (rr_0/\ve^2)^{\nu_n}\bigr]  \quad (r > r_0). \end{cases}
\end{equation}  

Knowing the Green's function, we evaluate the harmonic measure
density on the arc:
\begin{equation}
\omega(\theta|\x_0) = -(\partial_n G)_{r=R} = \frac{1}{\alpha R}\biggl(1 
+ 2\sum\limits_{n=1}^\infty \cos(\nu_n\theta)\cos(\nu_n\theta_0) (r_0/\ve)^{\nu_n}\biggr).
\end{equation}
Using the geometric series formula, one has
\begin{equation}
\sum\limits_{n=1}^\infty \cos(n\beta) \zeta^n = \zeta \, \frac{\cos\beta - \zeta}{1 - 2\zeta\cos\beta + \zeta^2} \,,
\end{equation}
so that
\begin{equation}
\omega(\theta|\x_0) = \frac{1}{\alpha R}\biggl(1 
+ \zeta \, \frac{\cos(\pi (\theta - \theta_0)/\alpha) - \zeta}{1 - 2\zeta \cos(\pi (\theta - \theta_0)/\alpha) + \zeta^2} 
+ \zeta \, \frac{\cos(\pi (\theta + \theta_0)/\alpha) - \zeta}{1 - 2\zeta \cos(\pi (\theta + \theta_0)/\alpha) + \zeta^2} \biggr), 
\end{equation}
where $\zeta = (r_0/\ve)^{\pi/\alpha}$.

In particular, we can compute the mean escape angle:
\begin{equation}
\E_{\x_0} \{ \theta_\tau \} = \ve \int\limits_0^\alpha \md\theta \, \theta \,\omega(\theta|\x_0)
= \frac{\alpha}{2} \biggl(1 + 4 \sum\limits_{n=1}^\infty \frac{(-1)^n - 1}{\pi^2 n^2} \cos(\nu_n \theta_0) (r_0/\ve)^{\nu_n}\biggr).
\end{equation}
For instance, if $\theta_0 = \alpha/2$, one gets $\E_{\x_0} \{
\theta_\tau \} = \alpha/2$ as expected.

\subsection{Mean acquired boundary local time}

Finally, we aim at evaluating the mean boundary local time $\ell_\tau$
acquired up to the escape moment $\tau$. Following
\cite{Grebenkov23}, we consider the joint probability density function
of $(\X_\tau, \ell_\tau,\tau)$, denoted as $j_D(\x,\ell,t|\x_0)$, with
$\x\in\pa_D$. On the one hand, the integral of this quantity over $t$ and
$\x\in\pa_D$ yields the (marginal) probability density function of
$\ell_\tau$:
\begin{equation} \label{eq:rhoD}
\rho_D(\ell|\x_0) = \int\limits_{\pa_D} \md\x \, \int\limits_0^\infty \md t \, j_D(\x,\ell,t|\x_0).
\end{equation}
On the other hand, multiplying this joint PDF by $e^{-q\ell}$ and
integrating over $\ell$ yields the probability flux density onto
$\pa_D$ in the presence of a partially reactive boundary $\Gamma$:
\begin{equation}
j_q(\x,t|\x_0) = \int\limits_0^\infty \md\ell \, e^{-q\ell}\, j_D(\x,\ell,t|\x_0).
\end{equation}
Its integral over $\x\in\pa_D$ yields the probability density of the
FPT to $\pa_D$ in the presence of a partially reactive boundary
$\Gamma$:
\begin{equation}
J_q(t|\x_0) = \int\limits_{\pa_D} \md\x \, \int\limits_0^\infty \md\ell \, e^{-q\ell}\, j_D(\x,\ell,t|\x_0).
\end{equation}
In particular, its integral over $t$ is the splitting probability,
i.e., the probability of hitting $\pa_D$ before reacting on $\Gamma$:
\begin{equation}
\tilde{J}_q(0|\x_0) = \int\limits_0^\infty \md t \, J_q(t|\x_0).
\end{equation}
We recall that the splitting probability $\tilde{J}_q(0|\x_0)$ satisfies:
\begin{equation}  \label{eq:splitting_eq}
\Delta \tilde{J}_q(0|\x_0) = 0, \qquad  \tilde{J}_q(0|\x_0) = 1 \quad \textrm{on}~\pa_D,
\qquad  (\partial_n + q) \tilde{J}_q(0|\x_0) = 0 \quad \textrm{on}~\Gamma.
\end{equation}
Comparing these expressions with Eq. (\ref{eq:rhoD}), we finally get
\begin{equation}
\tilde{J}_q(0|\x_0) = \int\limits_0^\infty \md\ell \, e^{-q\ell}\, \rho_D(\ell|\x_0) = \E_{\x_0} \{ e^{-q\ell_\tau}\} ,
\end{equation}
i.e., the splitting probability $\tilde{J}_q(0|\x_0)$ is the
moment-generating function of $\ell_\tau$. Note that this relation
could alternatively be deduced from spectral expansions derived in
\cite{Grebenkov23}, based on the Steklov-Dirichlet spectral problem. 
As a consequence,
$\tilde{J}_q(0|\x_0)$ determines all positive integer-order
moments of $\ell_\tau$. For instance, the mean acquired boundary
local time is
\begin{equation}
\E_{\x_0}\{ \ell_\tau\} = - \lim\limits_{q\to 0} \partial_q \tilde{J}_q(0|\x_0).
\end{equation}

In the limit $q\to 0$, we search $\tilde{J}_q(0|\x_0)$ as a formal
expansion:
\begin{equation}
\tilde{J}_q(0|\x_0) = v_0(\x_0) - q v_1(\x_0) + O(q^2).
\end{equation}
According to Eq. (\ref{eq:splitting_eq}), functions $v_0$ and $v_1$
satisfy
\begin{equation}
\Delta v_0 = 0, \qquad v_0|_{\pa_D} = 1, \qquad \partial_n v_0|_{\Gamma} = 0,
\end{equation}
and
\begin{equation}
\Delta v_1 = 0, \qquad v_1|_{\pa_D} = 0, \qquad \partial_n v_1|_{\Gamma} = v_0.
\end{equation}
One sees that $v_0 \equiv 1$, whereas $v_1$ can be found as
\begin{equation}
v_1(\x_0) = \int\limits_{\Gamma} \md\x \, G(\x,\x_0)\, v_0(\x),
\end{equation}
where $G(\x,\x_0)$ is the Green's function given by
Eq. (\ref{eq:Green_DN}). Evaluating this integral, we get
\begin{equation}
v_1(r_0,\theta_0) = 2\frac{\ve-r_0}{\alpha} + \frac{2\ve}{\alpha} \sum\limits_{n=1}^\infty \frac{(1 + (-1)^n)}{\nu_n^2-1} \cos(\nu_n\theta_0) 
\biggl(\frac{r_0}{\ve} - (r_0/\ve)^{\nu_n}\biggr).
\end{equation}  
We can therefore identify $v_1$ with $\E_{\x_0}\{ \ell_\tau\}$.  
Using the summation formula (see Table 2 from \cite{Grebenkov21}), we
can compute explicitly the first sum that yields
\begin{equation}
\E_{\x_0}\{ \ell_\tau\} = { \frac{2\ve}{\alpha} - r_0 \frac{\cos(\theta_0) + \cos(\alpha-\theta_0)}{\sin(\alpha)}  }
- \frac{2\ve}{\alpha} \sum\limits_{n=1}^\infty \frac{(1 + (-1)^n)}{\nu_n^2-1} \cos(\nu_n\theta_0) (r_0/\ve)^{\nu_n} .
\end{equation}  
We can simplify this expression as
\begin{equation}
\E_{\x_0}\{ \ell_\tau\} = \frac{2\ve}{\alpha} - r_0 \frac{\cos(\theta_0) + \cos(\alpha-\theta_0)}{\sin(\alpha)}  
- \frac{4\ve}{\alpha} \sum\limits_{n=1}^\infty \frac{\cos(2\nu_n\theta_0)}{4\nu_n^2-1}  (r_0/\ve)^{2\nu_n} .
\end{equation}
For instance, if the starting point $\x_0$ is located on the vertex
(i.e., $r_0 =0$), we get $\E_{0}\{\ell_\tau\} = 2\ve/\alpha$, i.e., it
increases as $\alpha$ decreases.

{ 

In the limit $\alpha \to 2\pi$, one needs to treat separately the
diverging contributions from the second term and the first term of the
sum. Setting $\alpha = 2\pi - \epsilon$ and evaluating the limit
$\epsilon\to 0$, we get for $\alpha = 2\pi$: 
\begin{equation}
\E_{\x_0}\{ \ell_\tau\} = \frac{\ve-r_0}{\pi} - r_0 \frac{\cos(\theta_0) \ln(r_0/\ve)}{\pi}  
+ \frac{2\ve}{\pi} \sum\limits_{n=2}^\infty \frac{\cos(n\theta_0)}{n^2-1} \biggl(r_0/\ve - (r_0/\ve)^n\biggr) .
\end{equation}  
Using the identity
\begin{equation}
\sum\limits_{n=2}^\infty \frac{\cos(n\theta_0)}{n^2-1} = \frac12 + \frac{\cos(\theta_0)}{4} - \sin(\theta_0) \frac{\pi - \theta_0}{2} \,,
\end{equation}
we get another representation for $\alpha = 2\pi$: 
\begin{equation}
\E_{\x_0}\{ \ell_\tau\} = \frac{\ve}{\pi} 
+ \frac{r_0}{\pi} \biggl(\frac{\cos(\theta_0)}{2} - (\pi-\theta_0)\sin(\theta_0) - \cos(\theta_0) \ln(r_0/\ve) \biggr) 
- \frac{2\ve}{\pi} \sum\limits_{n=2}^\infty \frac{\cos(n\theta_0)}{n^2-1} (r_0/\ve)^n .
\end{equation}  

}

\bibliographystyle{iopart-num-title}
\bibliography{EFW_jpa}

\end{document}